\documentclass[acmtog,authorversion]{acmart}

\usepackage{amsmath,amsfonts,bm}

\def\eqref#1{equation~\ref{#1}}

\def\1{\bm{1}}

\DeclareMathAlphabet{\mathsfit}{\encodingdefault}{\sfdefault}{m}{sl}
\SetMathAlphabet{\mathsfit}{bold}{\encodingdefault}{\sfdefault}{bx}{n}

\usepackage{multirow}
\usepackage{hyperref}
\usepackage{url}
\usepackage{algorithm}
\usepackage{algorithmic}
\newcommand{\Require}{\REQUIRE}
\newcommand{\For}{\FOR}
\newcommand{\EndFor}{\ENDFOR}
\newcommand{\State}{\STATE}

\usepackage[export]{adjustbox}
\usepackage{tikz}
\RequirePackage{luatex85,shellesc}
\usetikzlibrary{arrows.meta,positioning,calc}
\usepackage{pgfplots}
\usepackage{pgfplotstable}
\pgfplotsset{compat=1.17}

\definecolor{mygreen}{rgb}{0,0.8,0}
\definecolor{myred}{rgb}{0.8,0.0,0}

\definecolor{mygrey}{rgb}{0.40,0.40,0.40}    
\definecolor{mygrey2}{rgb}{0.60,0.60,0.60}   
\definecolor{mygreylight}{rgb}{0.85,0.85,0.85}    
\definecolor{mygreylighter}{rgb}{0.96,0.96,0.96}  
\definecolor{mygreylighterr}{rgb}{0.99,0.99,0.99} 

\copyrightyear{2023}
\acmYear{2023}
\setcopyright{acmlicensed}
\acmConference[SIGGRAPH '23 Conference Proceedings]{Special Interest Group on Computer Graphics and Interactive Techniques Conference Conference Proceedings}{August 6--10, 2023}{Los Angeles, CA, USA}
\acmBooktitle{Special Interest Group on Computer Graphics and Interactive Techniques Conference Conference Proceedings (SIGGRAPH '23 Conference Proceedings), August 6--10, 2023, Los Angeles, CA, USA}
\acmPrice{15.00}
\acmDOI{10.1145/3588432.3591540}
\acmISBN{979-8-4007-0159-7/23/08}

\begin{document}
\title{Iterative $\alpha$-(de)Blending: a Minimalist Deterministic Diffusion Model}
\author{Eric Heitz}
\orcid{0000-0002-9323-3318}
\affiliation{
  \institution{Unity Technologies}
  \country{France}
}
\author{Laurent Belcour}
\orcid{0000-0002-1982-0717}
\affiliation{
  \institution{Intel Corporation}
  \country{France}
}
\author{Thomas Chambon}
\orcid{0009-0003-2059-9778}
\affiliation{
  \institution{Unity Technologies}
  \country{France}
}

\begin{teaserfigure}
\vspace{-1mm}
\hspace{-10pt}
\begin{tikzpicture}
\begin{scope}
    \draw[->] (0, 1) -- (\linewidth-0.1\linewidth, 1);
    \node at (0.45\linewidth, 1.2) {increasing $\alpha$};
\end{scope}
\begin{scope}[shift={(0,0\linewidth)}]
    \node[inner sep=0, ]                (A0) { \fbox{\includegraphics[width=0.095\linewidth]{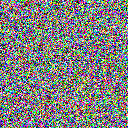}} };
    \node[inner sep=0, right=1pt of A0] (A1) { \fbox{\includegraphics[width=0.095\linewidth]{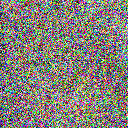}} };
    \node[inner sep=0, right=1pt of A1] (A2) { \fbox{\includegraphics[width=0.095\linewidth]{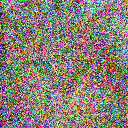}} };
    \node[inner sep=0, right=1pt of A2] (A3) { \fbox{\includegraphics[width=0.095\linewidth]{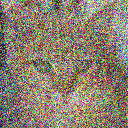}} };
    \node[inner sep=0, right=1pt of A3] (A4) { \fbox{\includegraphics[width=0.095\linewidth]{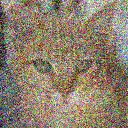}} };
    \node[inner sep=0, right=1pt of A4] (A5) { \fbox{\includegraphics[width=0.095\linewidth]{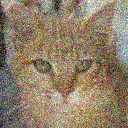}} };
    \node[inner sep=0, right=1pt of A5] (A6) { \fbox{\includegraphics[width=0.095\linewidth]{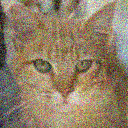}} };
    \node[inner sep=0, right=1pt of A6] (A7) { \fbox{\includegraphics[width=0.095\linewidth]{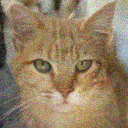}} };
    \node[inner sep=0, right=1pt of A7] (A8) { \fbox{\includegraphics[width=0.095\linewidth]{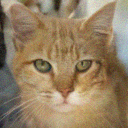}} };
    \node[inner sep=0, right=1pt of A8] (A9) { \fbox{\includegraphics[width=0.095\linewidth]{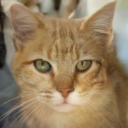}} };
    \node[rotate=90, left= 0pt of A0, anchor=south] {{ \footnotesize Noise}};
    \node[rotate=90, right=0pt of A9, anchor=north] {{ \footnotesize Cat}};
\end{scope}
\begin{scope}[shift={(2, -0.16\linewidth)}]
    \begin{scope}[shift={(0,0)}]
    \draw (-1, 0.0) node {...};
    \draw (0, 1.2) node {$\alpha_1$-\textbf{blend}};
    \draw (0, +0.0) node { \fbox{\includegraphics[width=0.07\linewidth]{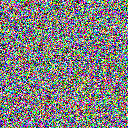}} };
    \draw (0, -1.0) node {$x_{\alpha_1} \sim p_{\alpha_1}$};
    \draw (0.8, 0.0) node {$\rightarrow$};
    \draw (1.9, 0) node { {\includegraphics[width=0.12\linewidth]{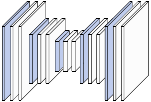}} };
    \draw (1.9, 1.2) node {$\alpha_1$-\textbf{deblend}};
    \draw (+3.0, +0.35) node {$\nearrow$};
    \draw (+3.0, -0.35) node {$\searrow$};
    \draw (+3.9, +0.7) node { \fbox{\includegraphics[width=0.07\linewidth]{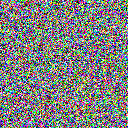}} };
    \draw (+3.9, -0.7) node { \fbox{\includegraphics[width=0.07\linewidth]{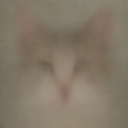}} };
    \draw (+3.9, +1.6) node {$\bar x_0$};
    \draw (+3.9, -1.6) node {$\bar x_1$};
    \draw (+4.8, +0.35) node {$\searrow$};
    \draw (+4.8, -0.35) node {$\nearrow$};
    \end{scope}
    \begin{scope}[shift={(5.8,0)}]
    \draw (0, 1.2) node {$\alpha_2$-\textbf{blend}};
    \draw (0, +0.0) node { \fbox{\includegraphics[width=0.07\linewidth]{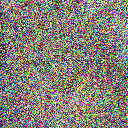}} };
    \draw (0, -1.0) node {$x_{\alpha_2} \sim p_{\alpha_2}$};
    \draw (0.8, 0.0) node {$\rightarrow$};
    \draw (1.9, 0) node { {\includegraphics[width=0.12\linewidth]{figures/teaser2/unet.pdf}} };
    \draw (1.9, 1.2) node {$\alpha_2$-\textbf{deblend}};
    \draw (+3.0, +0.35) node {$\nearrow$};
    \draw (+3.0, -0.35) node {$\searrow$};
    \draw (+3.9, +0.7) node { \fbox{\includegraphics[width=0.07\linewidth]{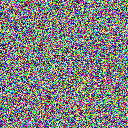}} };
    \draw (+3.9, -0.7) node { \fbox{\includegraphics[width=0.07\linewidth]{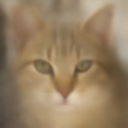}} };
    \draw (+3.9, +1.6) node {$\bar x_0$};
    \draw (+3.9, -1.6) node {$\bar x_1$};
    \draw (+4.8, +0.35) node {$\searrow$};
    \draw (+4.8, -0.35) node {$\nearrow$};
    \end{scope}
    \begin{scope}[shift={(2*5.8,0)}]
    \draw (0, 1.2) node {$\alpha_3$-\textbf{blend}};
    \draw (0, +0.0) node { \fbox{\includegraphics[width=0.07\linewidth]{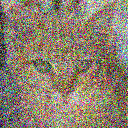}} };
    \draw (0, -1.0) node {$x_{\alpha_3} \sim p_{\alpha_3}$};
    \draw (1.0, 0.0) node {...};
    \end{scope}
\end{scope}
\end{tikzpicture}
\vspace{-7mm}
\caption{\label{fig:teaser} \textbf{Iterative $\alpha$-blending and deblending.} 
In this example, we map Gaussian noise to cat images.
We use a neural network trained to deblend blended Gaussian noise and cats. 
By deblending and reblending iteratively, we obtain a mapping between the Gaussian and cat densities.\\}
\vspace{-1mm}
\end{teaserfigure}

\begin{abstract}
\section*{Abstract}
\noindent We derive a minimalist but powerful deterministic denoising-diffusion model. While denoising diffusion has shown great success in many domains, its underlying theory remains largely inaccessible to non-expert users. Indeed, an understanding of graduate-level concepts such as Langevin dynamics or score matching appears to be required to grasp how it works. We propose an alternative approach that requires no more than undergrad calculus and probability. We consider two densities and observe what happens when random samples from these densities are blended (linearly interpolated). We show that iteratively blending and deblending samples produces random paths between the two densities that converge toward a deterministic mapping.
This mapping can be evaluated with a neural network trained to deblend samples. 
We obtain a model that behaves like deterministic denoising diffusion: it iteratively maps samples from one density (e.g., Gaussian noise) to another (e.g., cat images). 
However, compared to the state-of-the-art alternative, our model is simpler to derive, simpler to implement, more numerically stable, achieves higher quality results in our experiments, and has interesting connections to computer graphics. 
\end{abstract}

%
%


\settopmatter{
printacmref=false,
printccs=false,
}

\maketitle

\pagebreak
\section{Introduction}
\label{sec:introduction}

Diffusion models have recently become one of the most popular generative modeling tools~\citep{ramesh2022dalle2}.
They have outperformed state-of-the-art GANs~\citep{karras2019,Karras2021} and been applied to many applications, such as image generation~\citep{rombach2021latentdiffusion,dhariwal2021diffusion}, image processing~\citep{saharia2021superres,kawar2022restoration,whang2022deblurring}, text-to-image~\citep{saharia2022}, video~\citep{ho2022video} or audio~\citep{kong2020diffwave}.

\paragraph{First, there were stochastic diffusion models...}

These diffusion models can all be formulated as \textit{Stochastic Differential Equations} (SDEs) \cite{song2021sde} such as \textit{Langevin dynamics}.
Langevin's equation models a random walk that obeys a balance between two operations related to Gaussian noise: increasing noise by adding more noise, and decreasing noise by climbing the gradient of the log density. 
Increasing noise performs large steps but pushes the samples away from the true density. 
Decreasing noise projects the samples back onto the true density.
Carefully tracking and controlling this balance allows one to perform efficient random walks and provides a sampling procedure for the true density. 
This is the core of denoising diffusion approaches. 
Noise Conditional Score Networks (NCSNs)~\citep{song2019estimatinggradients,song2020a} use Langevin's equation directly by leveraging the fact that the score (the gradient of the log density in Langevin's equation) can be learnt via a denoiser when the samples are corrupted with Gaussian noise~\citep{vincent2011}.
Denoising Diffusion Probabilistic Models (DDPMs)~\citep{h02020ddpm,nichol2021improvedddpm} use a Markov chain formalism with a Gaussian prior that provides an SDE similar to Langevin dynamics, where the score is also implicitly learnt with a denoiser.

\paragraph{...then came deterministic diffusion models.}

Langevin's SDEs variants describe an equilibrium between noise injection and noise removal. 
Nullifying the noise injection in these SDEs yields \textit{Ordinary Differential Equations} (ODEs), also called \textit{Probability Flow ODEs}~\cite{song2021sde}, that simply describe the deterministic trajectory of a noisy sample projected back onto the true density.
For instance, Denoising Diffusion Implicit Models (DDIMs)~\citep{song2021ddim} are the ODE variants of DDPMs.
These ODEs provide a smooth, deterministic mapping between the Gaussian noise density and the true density.
Deterministic diffusion models have recently been proposed because an ODE requires far fewer solver iterations than its SDE counterpart. 
Furthermore, a deterministic mapping presents multiple practical advantages because samples are uniquely determined by their prior Gaussian noise.
For instance, they can be edited or interpolated via the Gaussian noise.

\paragraph{Is there a simpler approach to deterministic diffusion?}

The point of the above story is that, in the recent line of work on diffusion models, stochastic diffusion models came \textit{first} and deterministic diffusion models came \textit{after}, framed as special cases of the stochastic ones.
Hence they inherited the underlying mindset and mathematical framework.  
As a result, knowledge of advanced concepts such as Langevin dynamics, score matching, how they relate to Gaussian noise, etc., appears to be required to understand recent deterministic diffusion models.
We argue that this is an unnecessary detour for something that can be framed in a much simpler and more general way. 
We propose a fresh take on deterministic diffusion with another mindset, using only basic sampling concepts. \\

{
\setlength\leftmargini{4mm}
\begin{itemize}
\item \textbf{Simpler derivation.}
We derive a deterministic, diffusion-like model based on the sampling interpretation of blending and deblending. 
We call it Iterative $\alpha$-(de)Blending (IADB) in reference to the computer graphics \mbox{$\alpha$-blending} technique that composes images with a transparency parameter~\cite{porter1984}. 
Our model defines a mapping between arbitrary densities (of finite variance). \\

\item \textbf{Practical improvements.}
We show that, when the initial density is Gaussian, the mappings defined by IADB are exactly the same as the ones defined by DDIM~\citep{song2021ddim}, but with several benefits.
First, our derivation leads to a more numerically stable sampling formulation.
Second, our experiments show that IADB consistently outperforms DDIM in terms of final FID scores for several datasets and is more stable for a small number of sampling steps. \\

\item \textbf{Theoretical improvements.}
A side effect of our derivation is that, in contrast to DDIM, IADB does not require the assumption that the initial density is Gaussian, which is a significant generalization.
Furthermore, our derivation leads to a stochastic mapping algorithm that is reminiscent of computer graphics applications. 

\end{itemize}
}
\pagebreak
\section{Blending and deBlending as Sampling}
\label{sec:mapping_preliminaries}

\paragraph{Initial densities.}

\begin{figure}[h!]
\centering
\begin{tabular}{@{} c @{\hspace{1mm}} c @{}}
\textbf{$\alpha$-blending} & \textbf{$\alpha$-deblending} \\
$\underbrace{(x_0, x_1)}_{\sim p_0 \times p_1}
\rightarrow
\underbrace{x_\alpha}_{\sim p_\alpha}$ 
&
$\underbrace{x_\alpha}_{\sim p_\alpha}
\rightarrow
\underbrace{(x_0, x_1)}_{\sim p_0 \times p_1}$
\\
\scalebox{0.7}{\begin{tikzpicture}()
	\begin{scope}
		\clip (-1.0, -1.0) rectangle (5.0, 2.0);
		\draw[help lines] (-1.0, -1.0) grid (5.0, 2.0);
		\draw[fill=green,fill opacity=0.1,] (2.29, 0.71) -- (3.00, 0.00) -- (3.71, 0.71) -- (3.00, 1.41) -- (2.29, 0.71);
		\draw[fill=red!40!white,fill opacity=0.2,] (0.00, 1.00) -- (0.00, 0.00) -- (1.00, 0.00) -- (0.00, 1.00);
		\draw[] (0.47, 0.31) -- (3.25, 0.95);
		\fill[color=red!60!black,] (0.46571396526826336, 0.3056721711897118) circle (1.5pt);
		\fill[color=green!50!black,] (3.246667414374052, 0.953431600557339) circle (1.5pt);
		\fill[] (1.3, 0.5) circle (2pt);
		\draw (1.4,0.8) node {$x_\alpha$} ;
		\draw (3.0,-0.2) node {$x_1$} ;
		\draw (0.5,-0.2) node {$x_0$} ;
	\end{scope}
\end{tikzpicture}} &
\scalebox{0.7}{\begin{tikzpicture}()
	\begin{scope}
		\clip (-1.0, -1.0) rectangle (5.0, 2.0);
		\draw[help lines] (-1.0, -1.0) grid (5.0, 2.0);
		\draw[fill=green,fill opacity=0.1,] (2.29, 0.71) -- (3.00, 0.00) -- (3.71, 0.71) -- (3.00, 1.41) -- (2.29, 0.71);
		\draw[fill=red!40!white,fill opacity=0.2,] (0.00, 1.00) -- (0.00, 0.00) -- (1.00, 0.00) -- (0.00, 1.00);
            \fill[] (1.3, 0.5) circle (2pt);
		\draw[dotted] (0.47, 0.31) -- (3.25, 0.95);
		\fill[color=red!60!black,opacity=0.3] (0.46571396526826336, 0.3056721711897118) circle (1.5pt);
		\fill[color=green!50!black,opacity=0.3] (3.246667414374052, 0.953431600557339) circle (1.5pt);
            \draw[] (0.38703948055, 0.4980289591) -- (3.43024121205, 0.50459909543);
     	\fill[color=red!60!black] (0.38703948055, 0.4980289591) circle (1.5pt);
		\fill[color=green!50!black] (3.43024121205, 0.50459909543) circle (1.5pt);  
		\draw (3.0,-0.2) node {$x_1$} ;
		\draw (0.5,-0.2) node {$x_0$} ;
            \draw (1.4,0.8) node {$x_\alpha$} ;
	\end{scope}
\end{tikzpicture}} 
\end{tabular}
\vspace{-3mm}
\caption{\label{fig:basic_sampling_transformations_1} 
\textbf{Blending and deblending as sampling operations.}\vspace{-2mm}}
\end{figure}
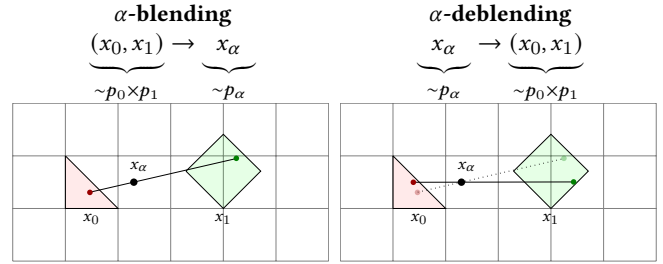

We consider two densities $p_0, p_1: \mathbb{R}^d \rightarrow \mathbb{R}^+$ represented, respectively, by the red triangle and the green square in Figure~\ref{fig:basic_sampling_transformations_1}.
We denote their corresponding samples as $x_0 \sim p_0$ and $x_1 \sim p_1$.
For independent samples $x_0$ and $x_1$, we use the notation $(x_0, x_1) \sim p_0 \times p_1$.

\paragraph{Definition of \textbf{$\alpha$-blending}.}

We use $p_\alpha$ to refer to the density of the blended samples \mbox{$x_\alpha = (1-\alpha) \, x_0 + \alpha \, x_1$} obtained by blending random samples $(x_0, x_1) \sim p_0 \times p_1$ with a parameter $\alpha \in [0,1]$.

\paragraph{Definition of \textbf{$\alpha$-deblending}.}

\begin{figure}[h!]
\centering
\begin{tabular}{@{\hspace{-1mm}} c @{\hspace{0.5mm}} c @{\hspace{0.5mm}} c @{}}
$\left(p_0 \times p_1\right)_{| (x_\alpha {\color{blue}\in \mathbb{R}^d}, \alpha)} \neq p_0 \times p_1$ 
&&
\\
\scalebox{0.7}{\begin{tikzpicture}()
	\begin{scope}
		\clip (-1.0, -1.0) rectangle (5.0, 2.0);
		\draw[help lines] (-1.0, -1.0) grid (5.0, 2.0);
		\draw[fill=green,fill opacity=0.1,] (2.29, 0.71) -- (3.00, 0.00) -- (3.71, 0.71) -- (3.00, 1.41) -- (2.29, 0.71);
		\draw[fill=red!40!white,fill opacity=0.2,] (0.00, 1.00) -- (0.00, 0.00) -- (1.00, 0.00) -- (0.00, 1.00);
		\draw[fill=green,fill opacity=0.1,opacity=0.1,] (0.35, 0.24) -- (0.05, 0.55) -- (-0.26, 0.24) -- (0.05, -0.06) -- (0.35, 0.24);
		\draw[fill=red!40!white,fill opacity=0.2,opacity=0.1,] (3.11, -1.06) -- (3.11, 1.27) -- (0.78, 1.27) -- (3.11, -1.06);
		\fill[] (0.9331254695867506, 0.38158289336002504) circle (2pt);
		\draw[] (0.02, 0.11) -- (3.07, 1.01);
		\draw[] (0.02, 0.19) -- (3.06, 0.82);
		\draw[] (0.07, 0.05) -- (2.95, 1.15);
		\draw[] (0.06, 0.32) -- (2.97, 0.51);
		\draw[] (0.23, 0.19) -- (2.58, 0.84);
		\fill[color=red!60!black,] (0.015326023210485866, 0.11127574894637196) circle (1.5pt);
		\fill[color=red!60!black,] (0.022195120429590043, 0.19279022272560173) circle (1.5pt);
		\fill[color=red!60!black,] (0.06743331095398836, 0.05100662104044394) circle (1.5pt);
		\fill[color=red!60!black,] (0.06201091613394106, 0.3249704858669217) circle (1.5pt);
		\fill[color=red!60!black,] (0.22676630234978776, 0.18665677031796732) circle (1.5pt);
		\fill[color=green!50!black,] (3.0746575111313676, 1.0122995636585488) circle (1.5pt);
		\fill[color=green!50!black,] (3.0586296176201246, 0.822099124840346) circle (1.5pt);
		\fill[color=green!50!black,] (2.953073839729862, 1.1529275287723808) circle (1.5pt);
		\fill[color=green!50!black,] (2.9657260943099724, 0.5136785108439329) circle (1.5pt);
		\fill[color=green!50!black,] (2.58129685980633, 0.8364105137914929) circle (1.5pt);
	\end{scope}
\end{tikzpicture}} & \raisebox{10mm}{$\searrow$} &
$\left(p_0 \times p_1\right)_{| (x_\alpha {\color{blue}\sim p_\alpha}, \alpha)} = p_0 \times p_1$ 
\\
\scalebox{0.7}{\begin{tikzpicture}()
	\begin{scope}
		\clip (-1.0, -1.0) rectangle (5.0, 2.0);
		\draw[help lines] (-1.0, -1.0) grid (5.0, 2.0);
		\draw[fill=green,fill opacity=0.1,] (2.29, 0.71) -- (3.00, 0.00) -- (3.71, 0.71) -- (3.00, 1.41) -- (2.29, 0.71);
		\draw[fill=red!40!white,fill opacity=0.2,] (0.00, 1.00) -- (0.00, 0.00) -- (1.00, 0.00) -- (0.00, 1.00);
		\draw[fill=green,fill opacity=0.1,opacity=0.1,] (1.07, -0.05) -- (0.77, 0.26) -- (0.47, -0.05) -- (0.77, -0.35) -- (1.07, -0.05);
		\draw[fill=red!40!white,fill opacity=0.2,opacity=0.1,] (4.80, -1.74) -- (4.80, 0.60) -- (2.46, 0.60) -- (4.80, -1.74);
		\fill[] (1.4385409830901308, 0.17881647070882098) circle (2pt);
		\draw[] (0.71, 0.09) -- (3.14, 0.38);
		\draw[] (0.66, 0.14) -- (3.27, 0.28);
		\draw[] (0.72, 0.16) -- (3.11, 0.22);
		\draw[] (0.69, 0.02) -- (3.19, 0.55);
		\draw[] (0.76, 0.05) -- (3.01, 0.49);
		\fill[color=red!60!black,] (0.7092203234493132, 0.0924210939509059) circle (1.5pt);
		\fill[color=red!60!black,] (0.6555032816690857, 0.13517517244188987) circle (1.5pt);
		\fill[color=red!60!black,] (0.721034559910283, 0.16325850035427938) circle (1.5pt);
		\fill[color=red!60!black,] (0.6874018159418619, 0.019584902265078873) circle (1.5pt);
		\fill[color=red!60!black,] (0.7631805307275424, 0.04704479094251634) circle (1.5pt);
		\fill[color=green!50!black,] (3.140289188918705, 0.3804056831439561) circle (1.5pt);
		\fill[color=green!50!black,] (3.265628953072569, 0.28064616666499353) circle (1.5pt);
		\fill[color=green!50!black,] (3.112722637176442, 0.2151184015360847) circle (1.5pt);
		\fill[color=green!50!black,] (3.1911990397694243, 0.5503567970775525) circle (1.5pt);
		\fill[color=green!50!black,] (3.014382038602837, 0.48628372349686505) circle (1.5pt);
	\end{scope}
\end{tikzpicture}} & \raisebox{10mm}{$\rightarrow$} &
\scalebox{0.7}{\input{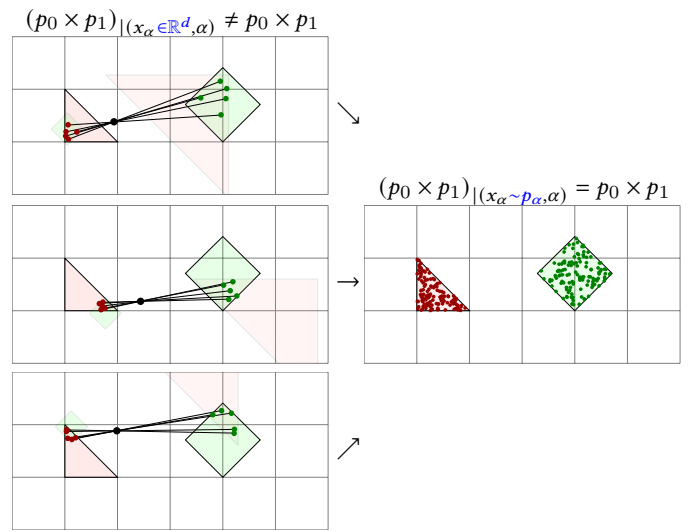}}
\\ 
\scalebox{0.7}{\begin{tikzpicture}()
	\begin{scope}
		\clip (-1.0, -1.0) rectangle (5.0, 2.0);
		\draw[help lines] (-1.0, -1.0) grid (5.0, 2.0);
		\draw[fill=green,fill opacity=0.1,] (2.29, 0.71) -- (3.00, 0.00) -- (3.71, 0.71) -- (3.00, 1.41) -- (2.29, 0.71);
		\draw[fill=red!40!white,fill opacity=0.2,] (0.00, 1.00) -- (0.00, 0.00) -- (1.00, 0.00) -- (0.00, 1.00);
		\draw[fill=green,fill opacity=0.1,opacity=0.1,] (0.43, 0.96) -- (0.12, 1.26) -- (-0.18, 0.96) -- (0.12, 0.66) -- (0.43, 0.96);
		\draw[fill=red!40!white,fill opacity=0.2,opacity=0.1,] (3.29, 0.61) -- (3.29, 2.94) -- (0.96, 2.94) -- (3.29, 0.61);
		\fill[] (0.9870714419126756, 0.8834222221580118) circle (2pt);
		\draw[] (0.05, 0.74) -- (3.16, 1.22);
		\draw[] (0.21, 0.75) -- (2.81, 1.19);
		\draw[] (0.13, 0.72) -- (2.98, 1.27);
		\draw[] (0.03, 0.90) -- (3.21, 0.84);
		\draw[] (0.03, 0.87) -- (3.22, 0.91);
		\fill[color=red!60!black,] (0.0541577079041568, 0.7403224333090713) circle (1.5pt);
		\fill[color=red!60!black,] (0.2056893502507879, 0.7523200637261568) circle (1.5pt);
		\fill[color=red!60!black,] (0.13466111905579814, 0.7195541677919077) circle (1.5pt);
		\fill[color=red!60!black,] (0.032301052388976466, 0.9040940195825004) circle (1.5pt);
		\fill[color=red!60!black,] (0.029630898013008257, 0.8711232846716412) circle (1.5pt);
		\fill[color=green!50!black,] (3.163870154599219, 1.2173217294722063) circle (1.5pt);
		\fill[color=green!50!black,] (2.81029632245708, 1.1893272584990071) circle (1.5pt);
		\fill[color=green!50!black,] (2.976028861912056, 1.265781015678921) circle (1.5pt);
		\fill[color=green!50!black,] (3.214869017467973, 0.8351880281675386) circle (1.5pt);
		\fill[color=green!50!black,] (3.2210993776785655, 0.912119742959543) circle (1.5pt);
	\end{scope}
\end{tikzpicture}} & \raisebox{10mm}{$\nearrow$}
\end{tabular}
\vspace{-3mm}
\caption{\label{fig:total_proba}\textbf{The law of total probability.}
Intuitively, deblending a \textbf{fixed} $x_\alpha~\in~\mathbb{R}^d$ means sampling only in a subset of the initial densities.
However, if $x_\alpha \sim p_\alpha$ is \textbf{random}, all these subsets are merged and the sampling occurs in the initial densities as if we had directly sampled $\left(x_0, x_1\right) \sim p_0 \times p_1$.}
\vspace{-10mm}
\end{figure}

We call the inverse sampling operation  $\alpha$-deblending, i.e., generating random $x_0$ and $x_1$ from the initial densities that could have been $\alpha$-blended to a point $x_\alpha$.
Formally, it means sampling random \textit{posteriors} \mbox{$\left(x_0, x_1\right)_{|(x_\alpha, \alpha)} \sim \left(p_0 \times p_1\right)_{| (x_\alpha, \alpha)}$}.
The key property is that if $x_\alpha \in \mathbb{R}^d$ is a \textbf{fixed} point, the posterior densities \textit{are not} the initial densities $p_0 \times p_1$.
However, if $x_\alpha \sim p_\alpha$ is a \textbf{random} sample, the posterior densities \textit{are} the initial densities.
This follows directly from the \emph{law of total probability} illustrated in Figure~\ref{fig:total_proba}.
In other words, $\alpha$-deblending a random sample $x_\alpha \sim p_\alpha$ is equivalent to sampling $\left(x_0, x_1\right) \sim p_0 \times p_1$.

\clearpage

\paragraph{Definition of \textbf{$\alpha$-(de)blending}.}

\begin{figure}[h!]
\centering
\begin{tabular}{@{} c @{}}
$\underbrace{\strut x_{\alpha_{1}}}_{\strut \sim p_{\alpha_1}} 
\rightarrow
\underbrace{\strut \left(x_0, x_1\right)}_{\strut \sim p_0 \times p_1} 
\rightarrow
\underbrace{\strut x_{\alpha_{2}}}_{\strut \sim p_{\alpha_2}}$
\\
\begin{tikzpicture}()
	\begin{scope}
		\clip (-1.0, -1.0) rectangle (5.0, 2.0);
		\draw[help lines] (-1.0, -1.0) grid (5.0, 2.0);
		\draw[fill=green,fill opacity=0.1,] (2.29, 0.71) -- (3.00, 0.00) -- (3.71, 0.71) -- (3.00, 1.41) -- (2.29, 0.71);
		\draw[fill=red!40!white,fill opacity=0.2,] (0.00, 1.00) -- (0.00, 0.00) -- (1.00, 0.00) -- (0.00, 1.00);
		\draw[] (0.45, 0.47) -- (3.27, 0.57);
		\fill[color=red!60!black,] (0.45414431755085094, 0.46800732716294535) circle (1.5pt);
		\fill[color=green!50!black,] (3.2736632590480146, 0.5746495699531274) circle (1.5pt);
		\draw[arrows = {-Stealth[reversed, reversed]},color=gray,] (1.30, 0.50) to [bend left=45] (1.86, 0.52);
		\fill[] (1.3, 0.5) circle (2pt);
		\fill[] (1.863903788299433, 0.5213284485580364) circle (2pt);
		\draw (1.25,0.83) node {$x_{\alpha_1}$};
		\draw (1.80,0.83) node {$x_{\alpha_2}$};
		\draw (3.0,-0.2) node {$x_1$} ;
		\draw (0.5,-0.2) node {$x_0$} ;
	\end{scope}
\end{tikzpicture}
\end{tabular}
\vspace{-3mm}
\caption{\label{fig:basic_sampling_transformations_3} 
\textbf{$\alpha$-(de)blending.}} 
\end{figure}
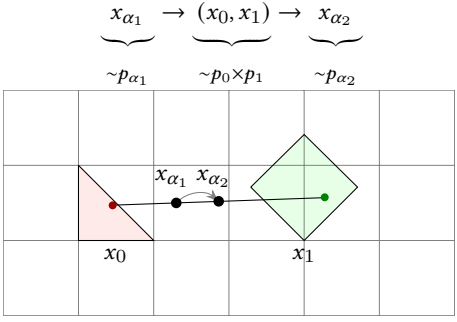

Let's consider two blending parameters $\alpha_1, \alpha_2 \in [0,1]$.
Using the previous proposition, we can chain $\alpha_1$-deblending and $\alpha_2$-blending to map a random sample $x_{\alpha_1} \sim p_{\alpha_1}$ to a random sample $x_{\alpha_2} \sim p_{\alpha_2}$.
Indeed, by sampling posteriors for a random sample $x_{\alpha_1} \sim p_{\alpha_1}$, we obtain random samples $\left(x_0, x_1\right) \sim \left(p_0 \times p_1\right)$ from the initial densities, and blending them with parameter $\alpha_2$ provides a random sample $x_{\alpha_2} \sim p_{\alpha_2}$. 
This is illustrated in Figure~\ref{fig:basic_sampling_transformations_3}.

\section{Iterative $\alpha$-(de)Blending (IADB)}
\label{sec:iterative_algorithms}

\begin{figure}[h!]
\centering
\begin{tabular}{@{} c @{}}
$\underbrace{x_0}_{\sim p_0}  \rightarrow ..\rightarrow \underbrace{x_\alpha}_{\sim p_\alpha}  \rightarrow .. \rightarrow \underbrace{x_1}_{\sim p_1}$ 
\\
\input{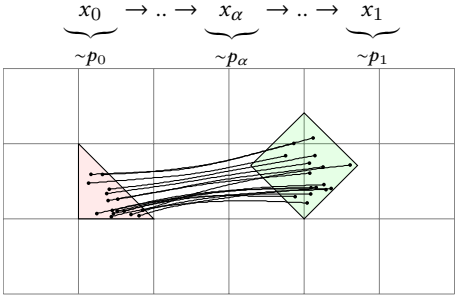}
\end{tabular}
\vspace{-3mm}
\caption{\label{fig:basic_sampling_transformations_4} 
\textbf{Iterative $\alpha$-(de)blending}} 
\end{figure}

Our objective is to define a deterministic mapping such that i.i.d. samples $x_0 \sim p_0$ passed through the mapping produce i.i.d. samples $x_1 \sim p_1$. 
We introduce Iterative $\alpha$-(de)Blending (IADB), an iterative algorithm that can be implemented stochastically or deterministically. 
Our main result is that both variants converge toward the same limit, which yields a deterministic mapping between the densities $p_0$ and $p_1$, as shown in Figure~\ref{fig:basic_sampling_transformations_4}.

\paragraph{Algorithm~\ref{algo:iterative_stochastic_sampling}: iterative $\alpha$-(de)blending ({\color{blue}stochastic}).} 

Let's consider a number of iterations $T$ and evenly distributed blending parameters $\alpha_t = t/T, t=\{0,..,T\})$.
This algorithm creates a sequence $(x_{\alpha_t} \sim p_{\alpha_t}, t=\{0,..,T\})$ that starts with a random sample $x_0 \sim p_0$ and ends with a random sample $x_{\alpha_T} = x_1 \sim p_1$ by applying $\alpha$-(de)blending iteratively. In each iteration, $x_{\alpha_t} \sim p_{\alpha_t}$ is $\alpha_t$-deblended by sampling random posteriors, which are sampled and $\alpha_{t+1}$-blended again to obtain a new sample $x_{\alpha_{t+1}} \sim p_{\alpha_{t+1}}$.
End to end, this algorithm provides a stochastic mapping between samples $x_0 \sim p_0$ and samples~$x_1 \sim p_1$.

\begin{algorithm}[H]
\centering
\caption{Iterative $\alpha$-(de)blending ({\color{blue}stochastic})} \label{algo:iterative_stochastic_sampling}
\begin{algorithmic}
\Require $x_0 \sim p_0$, $T$, $\alpha_t := \frac{t}{T}$
\For{$t=0,..,T-1$}
\State {\color{blue}sample $\left({x}_{0}, {x}_{1}\right) \sim \left(p_{0} \times p_{1}\right)_{|(x_{\alpha_t}, \alpha_t)}$}
\State $x_{\alpha_{t+1}} = (1-\alpha_{t+1})\,x_0 + \alpha_{t+1}\,x_1$
\EndFor
\end{algorithmic}
\end{algorithm}

\paragraph{Algorithm~\ref{algo:iterative_deterministic_sampling}: iterative $\alpha$-(de)blending ({\color{blue}deterministic}).} 

This algorithm is the same as Algorithm~\ref{algo:iterative_stochastic_sampling} except that, in each iteration, the random posterior samples are replaced by their expectations. The algorithm is thus not stochastic but deterministic.

\begin{algorithm}[H]
\centering
\caption{Iterative $\alpha$-(de)blending ({\color{blue}deterministic})} \label{algo:iterative_deterministic_sampling}
\begin{algorithmic}
\Require $x_0 \sim p_0$, $T$, $\alpha_t := \frac{t}{T}$
\For{$t=0,..,T-1$}
\State {\color{blue}$(\bar{x}_0, \bar{x}_1) = \mathop{\mathbb{E}}_{\left(p_{0} \times p_{1}\right)_{|(x_{\alpha_t}, \alpha_t)}}\left[\left({x}_{0}, {x}_{1}\right) \right]$}
\State $x_{\alpha_{t+1}} = (1-\alpha_{t+1})\,\bar{x}_0 + \alpha_{t+1}\,\bar{x}_1$
\EndFor
\end{algorithmic}
\end{algorithm}

\paragraph{\normalfont\textbf{Theorem:} \textit{convergence of iterative $\alpha$-(de)blending.}}

If $p_0$ and $p_1$ are Riemann-integrable densities of finite variance, the sequences computed by 
Algorithm~\ref{algo:iterative_stochastic_sampling} and Algorithm~\ref{algo:iterative_deterministic_sampling} converge toward the same limit as the number of steps $T$ increases, i.e., for any $\alpha \in [0,1]$:
\begin{align}
\lim_{T \rightarrow \infty} &x_{\alpha} \text{ computed by Algorithm~\ref{algo:iterative_stochastic_sampling}}(x_0, T) \nonumber\\
&=
\lim_{T \rightarrow \infty} x_{\alpha} \text{ computed by Algorithm~\ref{algo:iterative_deterministic_sampling}}(x_0, T).
\end{align}

\paragraph{\normalfont\textbf{Proof.}}

\begin{figure}[h!]
\centering
\begin{tabular}{@{} c @{\hspace{0.5mm}} c @{\hspace{0.5mm}} c @{\hspace{0.5mm}} c @{}}
& $T=2$ steps & $T=10$ steps & $T=1000$ steps \\
\raisebox{3mm}{\rotatebox{90}{Alg.~\ref{algo:iterative_stochastic_sampling}}} &
\scalebox{0.44}{\begin{tikzpicture}()
	\begin{scope}
		\clip (-1.0, -1.0) rectangle (5.0, 2.0);
		\draw[help lines] (-1.0, -1.0) grid (5.0, 2.0);
		\draw[fill=green,fill opacity=0.1,] (2.29, 0.71) -- (3.00, 0.00) -- (3.71, 0.71) -- (3.00, 1.41) -- (2.29, 0.71);
		\draw[fill=red!40!white,fill opacity=0.2,] (0.00, 1.00) -- (0.00, 0.00) -- (1.00, 0.00) -- (0.00, 1.00);
		\draw[color=red,opacity=0.5,] (0.10, 0.80) -- (1.30, 0.86) -- (2.59, 0.89);
		\fill[color=red,opacity=0.5,] (0.1, 0.8) circle (0.7pt);
		\fill[color=red,opacity=0.5,] (2.59023284099254, 0.8859270711168936) circle (0.7pt);
		\draw[color=red,opacity=0.5,] (0.10, 0.80) -- (1.54, 1.06) -- (3.03, 1.37);
		\fill[color=red,opacity=0.5,] (0.1, 0.8) circle (0.7pt);
		\fill[color=red,opacity=0.5,] (3.0331592045577076, 1.3664724897163727) circle (0.7pt);
		\draw[color=red,opacity=0.5,] (0.10, 0.80) -- (1.41, 0.73) -- (2.76, 1.17);
		\fill[color=red,opacity=0.5,] (0.1, 0.8) circle (0.7pt);
		\fill[color=red,opacity=0.5,] (2.7583428345083463, 1.1717833669709146) circle (0.7pt);
		\draw[color=red,opacity=0.5,] (0.10, 0.80) -- (1.42, 0.58) -- (2.76, 0.39);
		\fill[color=red,opacity=0.5,] (0.1, 0.8) circle (0.7pt);
		\fill[color=red,opacity=0.5,] (2.755597625991783, 0.3905304177741873) circle (0.7pt);
		\draw[color=red,opacity=0.5,] (0.10, 0.80) -- (1.35, 0.77) -- (2.69, 0.86);
		\fill[color=red,opacity=0.5,] (0.1, 0.8) circle (0.7pt);
		\fill[color=red,opacity=0.5,] (2.686711240478021, 0.8637794764417615) circle (0.7pt);
		\draw[color=red,opacity=0.5,] (0.10, 0.80) -- (1.49, 1.01) -- (2.82, 1.22);
		\fill[color=red,opacity=0.5,] (0.1, 0.8) circle (0.7pt);
		\fill[color=red,opacity=0.5,] (2.817735429771468, 1.2213898114490154) circle (0.7pt);
		\draw[color=red,opacity=0.5,] (0.10, 0.80) -- (1.30, 0.84) -- (2.50, 0.91);
		\fill[color=red,opacity=0.5,] (0.1, 0.8) circle (0.7pt);
		\fill[color=red,opacity=0.5,] (2.5018473241687813, 0.9057526852498724) circle (0.7pt);
		\draw[color=red,opacity=0.5,] (0.10, 0.80) -- (1.60, 0.46) -- (2.70, 0.62);
		\fill[color=red,opacity=0.5,] (0.1, 0.8) circle (0.7pt);
		\fill[color=red,opacity=0.5,] (2.6962189388855995, 0.6196527502171063) circle (0.7pt);
		\draw[color=red,opacity=0.5,] (0.10, 0.80) -- (1.74, 0.70) -- (3.41, 0.72);
		\fill[color=red,opacity=0.5,] (0.1, 0.8) circle (0.7pt);
		\fill[color=red,opacity=0.5,] (3.4136802553598837, 0.7212013303390247) circle (0.7pt);
		\draw[color=red,opacity=0.5,] (0.10, 0.80) -- (1.41, 0.66) -- (2.56, 0.81);
		\fill[color=red,opacity=0.5,] (0.1, 0.8) circle (0.7pt);
		\fill[color=red,opacity=0.5,] (2.5581951043633198, 0.8084112879669278) circle (0.7pt);
	\end{scope}
\end{tikzpicture}} &
\scalebox{0.44}{\begin{tikzpicture}()
	\begin{scope}
		\clip (-1.0, -1.0) rectangle (5.0, 2.0);
		\draw[help lines] (-1.0, -1.0) grid (5.0, 2.0);
		\draw[fill=green,fill opacity=0.1,] (2.29, 0.71) -- (3.00, 0.00) -- (3.71, 0.71) -- (3.00, 1.41) -- (2.29, 0.71);
		\draw[fill=red!40!white,fill opacity=0.2,] (0.00, 1.00) -- (0.00, 0.00) -- (1.00, 0.00) -- (0.00, 1.00);
		\draw[color=red,opacity=0.5,] (0.10, 0.80) -- (0.37, 0.81) -- (0.70, 0.73) -- (1.05, 0.68) -- (1.33, 0.74) -- (1.66, 0.85) -- (1.97, 0.94) -- (2.30, 1.06) -- (2.62, 1.18) -- (2.95, 1.25);
		\fill[color=red,opacity=0.5,] (0.1, 0.8) circle (0.7pt);
		\fill[color=red,opacity=0.5,] (2.947684707623623, 1.2504065662755313) circle (0.7pt);
		\draw[color=red,opacity=0.5,] (0.10, 0.80) -- (0.46, 0.78) -- (0.78, 0.85) -- (1.17, 0.84) -- (1.48, 0.88) -- (1.81, 0.90) -- (2.13, 1.00) -- (2.43, 1.12) -- (2.77, 1.18) -- (3.09, 1.26);
		\fill[color=red,opacity=0.5,] (0.1, 0.8) circle (0.7pt);
		\fill[color=red,opacity=0.5,] (3.086200300312562, 1.2577222031741189) circle (0.7pt);
		\draw[color=red,opacity=0.5,] (0.10, 0.80) -- (0.48, 0.77) -- (0.78, 0.83) -- (1.08, 0.85) -- (1.42, 0.83) -- (1.76, 0.89) -- (2.08, 0.90) -- (2.39, 1.04) -- (2.72, 1.13) -- (3.04, 1.19);
		\fill[color=red,opacity=0.5,] (0.1, 0.8) circle (0.7pt);
		\fill[color=red,opacity=0.5,] (3.043948823658271, 1.1939138634526199) circle (0.7pt);
		\draw[color=red,opacity=0.5,] (0.10, 0.80) -- (0.44, 0.78) -- (0.71, 0.74) -- (1.03, 0.68) -- (1.36, 0.74) -- (1.64, 0.83) -- (1.95, 0.95) -- (2.25, 1.00) -- (2.55, 1.05) -- (2.87, 1.14);
		\fill[color=red,opacity=0.5,] (0.1, 0.8) circle (0.7pt);
		\fill[color=red,opacity=0.5,] (2.865782450731802, 1.142077545547139) circle (0.7pt);
		\draw[color=red,opacity=0.5,] (0.10, 0.80) -- (0.38, 0.76) -- (0.65, 0.77) -- (0.94, 0.77) -- (1.21, 0.75) -- (1.50, 0.74) -- (1.80, 0.76) -- (2.08, 0.86) -- (2.37, 0.85) -- (2.65, 0.88);
		\fill[color=red,opacity=0.5,] (0.1, 0.8) circle (0.7pt);
		\fill[color=red,opacity=0.5,] (2.6511497573843417, 0.8780504411628708) circle (0.7pt);
		\draw[color=red,opacity=0.5,] (0.10, 0.80) -- (0.42, 0.71) -- (0.72, 0.74) -- (1.02, 0.71) -- (1.36, 0.74) -- (1.68, 0.71) -- (1.95, 0.76) -- (2.21, 0.87) -- (2.50, 0.95) -- (2.80, 0.99);
		\fill[color=red,opacity=0.5,] (0.1, 0.8) circle (0.7pt);
		\fill[color=red,opacity=0.5,] (2.8020474782133973, 0.9899817115295233) circle (0.7pt);
		\draw[color=red,opacity=0.5,] (0.10, 0.80) -- (0.41, 0.85) -- (0.70, 0.83) -- (1.02, 0.89) -- (1.35, 0.97) -- (1.65, 1.00) -- (1.96, 1.06) -- (2.27, 1.13) -- (2.59, 1.17) -- (2.91, 1.25);
		\fill[color=red,opacity=0.5,] (0.1, 0.8) circle (0.7pt);
		\fill[color=red,opacity=0.5,] (2.910125234884937, 1.245910529339161) circle (0.7pt);
		\draw[color=red,opacity=0.5,] (0.10, 0.80) -- (0.42, 0.74) -- (0.80, 0.74) -- (1.17, 0.74) -- (1.55, 0.77) -- (1.93, 0.77) -- (2.27, 0.78) -- (2.58, 0.83) -- (2.83, 0.94) -- (3.17, 1.01);
		\fill[color=red,opacity=0.5,] (0.1, 0.8) circle (0.7pt);
		\fill[color=red,opacity=0.5,] (3.17430639683165, 1.0121285897606345) circle (0.7pt);
		\draw[color=red,opacity=0.5,] (0.10, 0.80) -- (0.44, 0.74) -- (0.75, 0.67) -- (1.06, 0.75) -- (1.34, 0.75) -- (1.65, 0.84) -- (1.97, 0.84) -- (2.25, 0.95) -- (2.52, 1.05) -- (2.83, 1.16);
		\fill[color=red,opacity=0.5,] (0.1, 0.8) circle (0.7pt);
		\fill[color=red,opacity=0.5,] (2.826630184072594, 1.1604332821623002) circle (0.7pt);
		\draw[color=red,opacity=0.5,] (0.10, 0.80) -- (0.44, 0.80) -- (0.83, 0.79) -- (1.13, 0.81) -- (1.49, 0.78) -- (1.85, 0.83) -- (2.19, 0.92) -- (2.49, 1.02) -- (2.77, 1.11) -- (3.07, 1.18);
		\fill[color=red,opacity=0.5,] (0.1, 0.8) circle (0.7pt);
		\fill[color=red,opacity=0.5,] (3.0742211295907547, 1.183003286853369) circle (0.7pt);
	\end{scope}
\end{tikzpicture}} &
\scalebox{0.44}{\input{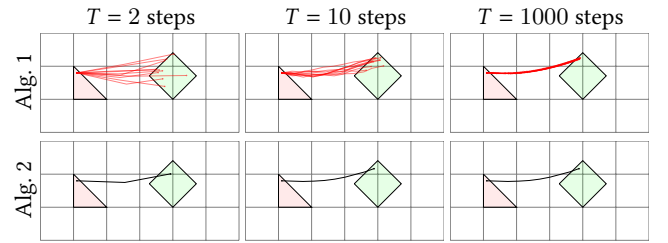}} \\
\raisebox{3mm}{\rotatebox{90}{Alg.~\ref{algo:iterative_deterministic_sampling}}} &
\scalebox{0.44}{\begin{tikzpicture}()
	\begin{scope}
		\clip (-1.0, -1.0) rectangle (5.0, 2.0);
		\draw[help lines] (-1.0, -1.0) grid (5.0, 2.0);
		\draw[fill=green,fill opacity=0.1,] (2.29, 0.71) -- (3.00, 0.00) -- (3.71, 0.71) -- (3.00, 1.41) -- (2.29, 0.71);
		\draw[fill=red!40!white,fill opacity=0.2,] (0.00, 1.00) -- (0.00, 0.00) -- (1.00, 0.00) -- (0.00, 1.00);
		\draw[color=black,] (0.10, 0.80) -- (1.55, 0.75) -- (2.93, 1.01);
		\fill[] (0.1, 0.8) circle (0.7pt);
		\fill[] (2.9319789955727917, 1.0082556389202169) circle (0.7pt);
	\end{scope}
\end{tikzpicture}} &
\scalebox{0.44}{\begin{tikzpicture}()
	\begin{scope}
		\clip (-1.0, -1.0) rectangle (5.0, 2.0);
		\draw[help lines] (-1.0, -1.0) grid (5.0, 2.0);
		\draw[fill=green,fill opacity=0.1,] (2.29, 0.71) -- (3.00, 0.00) -- (3.71, 0.71) -- (3.00, 1.41) -- (2.29, 0.71);
		\draw[fill=red!40!white,fill opacity=0.2,] (0.00, 1.00) -- (0.00, 0.00) -- (1.00, 0.00) -- (0.00, 1.00);
		\draw[color=black,] (0.10, 0.80) -- (0.42, 0.79) -- (0.74, 0.78) -- (1.07, 0.79) -- (1.39, 0.81) -- (1.70, 0.85) -- (2.01, 0.91) -- (2.31, 0.99) -- (2.61, 1.07) -- (2.89, 1.17);
		\fill[] (0.1, 0.8) circle (0.7pt);
		\fill[] (2.8880715142007394, 1.1655022449545334) circle (0.7pt);
	\end{scope}
\end{tikzpicture}} &
\scalebox{0.44}{\begin{tikzpicture}()
	\begin{scope}
		\clip (-1.0, -1.0) rectangle (5.0, 2.0);
		\draw[help lines] (-1.0, -1.0) grid (5.0, 2.0);
		\draw[fill=green,fill opacity=0.1,] (2.29, 0.71) -- (3.00, 0.00) -- (3.71, 0.71) -- (3.00, 1.41) -- (2.29, 0.71);
		\draw[fill=red!40!white,fill opacity=0.2,] (0.00, 1.00) -- (0.00, 0.00) -- (1.00, 0.00) -- (0.00, 1.00);
		\draw[color=black,] (0.10, 0.80) -- (0.42, 0.79) -- (0.74, 0.78) -- (1.07, 0.79) -- (1.39, 0.81) -- (1.70, 0.85) -- (2.01, 0.91) -- (2.31, 0.99) -- (2.61, 1.07) -- (2.89, 1.17);
		\fill[] (0.1, 0.8) circle (0.7pt);
		\fill[] (2.8880715142007394, 1.1655022449545334) circle (0.7pt);
	\end{scope}
\end{tikzpicture}}
\end{tabular}
\vspace{-3mm}
\caption{\label{fig:algorithms}
Both algorithms step iteratively by moving the samples along segments defined by their posterior densities. 
The difference is that Algorithm~\ref{algo:iterative_stochastic_sampling} uses segments between random posterior samples, which creates stochastic paths, while Algorithm~\ref{algo:iterative_deterministic_sampling} uses the segment between the average of the posterior samples, which creates deterministic paths. 
As the number of steps $T$ increases, the randomness of the stochastic paths averages out and they converge toward the deterministic paths.
}
\end{figure}

We provide a detailed proof in Appendix~A of our supplemental.
But intuitively, with each iteration, Algorithm~\ref{algo:iterative_stochastic_sampling} makes a small step $\Delta x_\alpha = (x_1-x_0)\, \Delta\alpha$ along the segment given by random posterior samples. 
As the number of iterations increases, many small random steps average out, and the infinitesimal steps are described by an ODE that involves the expected posteriors, as in Algorithm~\ref{algo:iterative_deterministic_sampling}:
\begin{align}
\mathrm{d}x_\alpha = \left( \bar{x}_{1} - \bar{x}_{0}\right) \, \mathrm{d}\alpha. 
\label{eq:ODE}
\end{align}
Hence, as $T$ increases, the update rule of Algorithm~\ref{algo:iterative_stochastic_sampling} converges toward that of Algorithm~\ref{algo:iterative_deterministic_sampling}.
This is shown in Figure~\ref{fig:algorithms}.

\pagebreak

\paragraph{Connection to computer graphics applications.}

\begin{figure}[h!]
\centering
\begin{tabular}{@{} c @{\hspace{2mm}} c @{\hspace{0.5mm}} c @{\hspace{0.5mm}} c @{}}
$x_0 \sim p_0$ & 
$x_0 \xrightarrow[T=32]{\text{Alg.~\ref{algo:iterative_stochastic_sampling}}} x_1$ & 
$x_0 \xrightarrow[T=512]{\text{Alg.~\ref{algo:iterative_stochastic_sampling}}} x_1$ & 
$x_0 \xrightarrow[T=1M]{\text{Alg.~\ref{algo:iterative_stochastic_sampling}}} x_1$ \\
\includegraphics[width=0.24\linewidth]{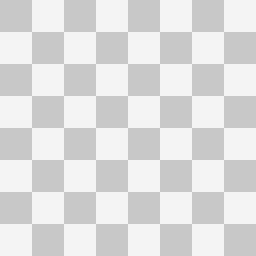} &
\includegraphics[width=0.24\linewidth]{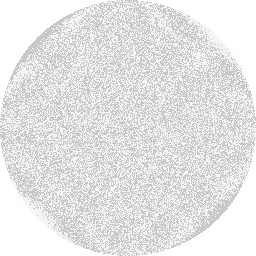} &
\includegraphics[width=0.24\linewidth]{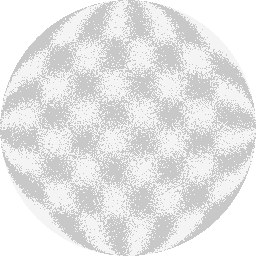} &
\includegraphics[width=0.24\linewidth]{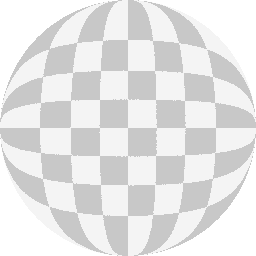}
\end{tabular}
\vspace{-3mm}
\caption{\label{fig:param}
\textbf{Stochastic mapping with Algorithm~\ref{algo:iterative_stochastic_sampling}.}
We map a uniform density on a square ($p_0$) and to a uniform density on a disk ($p_1$). 
The checkerboard pattern shows the randomness of the resulting mapping.
The larger the number of steps $T$, the more the mapping converges and reveals a smooth parameterization of the disk.
}
\end{figure}

\begin{figure}[h!]
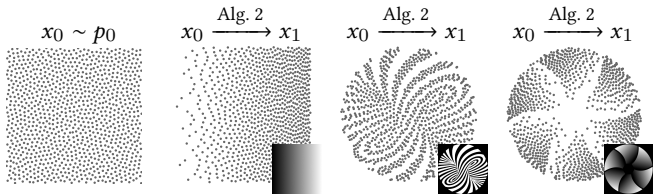

\centering
\begin{tabular}{@{} c @{\hspace{3mm}} c @{\hspace{-0.75mm}} c @{\hspace{-0.75mm}} c @{}}
$x_0 \sim p_0$ & 
\hspace{-5mm}$x_0 \xrightarrow{\text{Alg.~\ref{algo:iterative_deterministic_sampling}}} x_1$ & 
\hspace{-5mm}$x_0 \xrightarrow{\text{Alg.~\ref{algo:iterative_deterministic_sampling}}} x_1$ & 
\hspace{-5mm}$x_0 \xrightarrow{\text{Alg.~\ref{algo:iterative_deterministic_sampling}}} x_1$ \\
\raisebox{3.2mm}{\scalebox{1.8}{\input{figures/bluenoise/samples.tex}}} &
\scalebox{1.8}{\input{figures/bluenoise/ramp.tex}} &
\scalebox{1.8}{\input{figures/bluenoise/vortex.tex}} &
\scalebox{1.8}{\input{figures/bluenoise/shutter2.tex}}
\end{tabular}
\vspace{-3mm}
\caption{\label{fig:bluenoise}
\textbf{Deterministic mapping with Algorithm~\ref{algo:iterative_deterministic_sampling}.}
We use the deterministic mapping to warp blue-noise samples in the unit square to arbitrary densities.
}
\end{figure}

Figures~\ref{fig:param} and \ref{fig:bluenoise} show how the mapping behaves in 2D.
The deterministic mapping defined by the limit of the algorithm is a transport map (also called an area-preserving parameterization) that could potentially be of interest for common computer graphics applications such as parameterizing, sampling, and stippling. 
We believe that showing the connection is interesting, but our point here is not to make competitive claims for these applications. Instead, our focus is on using this mapping for deterministic denoising diffusion, as presented in Section~\ref{sec:learning}.

\section{Learning Iterative $\alpha$-(de)Blending}
\label{sec:learning}

In this section, we explain how to use iterative $\alpha$-(de)blending in a machine learning context, where we train a neural network $D_\theta$ to predict the average posterior samples used in Algorithm~\ref{algo:iterative_deterministic_sampling}. 

\subsection{Variant Formulations of Iterative $\alpha$-(de)Blending}
\label{sec:variant}

A direct transposition of Algorithm~\ref{algo:iterative_deterministic_sampling} means learning the averages of both posterior samples $\bar{x}_0$ and $\bar{x}_1$.
However, one is implicitly given by the other such that it is not necessary to learn both, and variants of Alg.~\ref{algo:iterative_deterministic_sampling} are possible.
The fact that multiple, theoretically equivalent, variants are possible is pointed out by~\citet{salimans2022progressive}.
However, they are not equivalent in practice. 
In Table~\ref{tab:equivalent_formulations}, we summarize four variants derived in Appendix~B of our supplemental and compare their practical properties.
Variant (a) is the vanilla transposition of Algorithm~\ref{algo:iterative_deterministic_sampling}.
It is highly unstable because instead of being a numerical update of the current sample $x_{\alpha_t}$, the new sample $x_{\alpha_{t+1}}$ is computed from the outputs of the neural network.
The residual learning errors of the network accumulate at each step and the larger the number of steps $T$, the more this variant diverges.
Variants (b) and (c) consist of learning either only $\bar{x}_0$ or $\bar{x}_1$. 
The sampling suffers from numerical instability near $\alpha_t=0$ and $\alpha_t=1$ because of the respective divisions by $\alpha_t$ and $1-\alpha_{t}$.
We recommend using variant (d), which consists of learning the average difference vector $\bar{x}_1 - \bar{x}_0$.
It is a direct transposition of the ODE defined in Equation~\ref{eq:ODE}.
This variant updates the current samples at each iteration without any division, making it the most stable variant for both training and sampling.

\begin{table}[!h]
\centering
\scalebox{0.66}{
\begin{tabular}{@{} c @{}}
\begin{tabular}{@{} | c | c | c | c | @{}}
\hline
(a) \textbf{learn $\bar{x}_0$ and $\bar{x}_1$} &
(b) \textbf{learn only $\bar{x}_0$} & 
(c) \textbf{learn only $\bar{x}_1$} & 
(d) \textbf{learn $\bar{x}_1 - \bar{x}_0$} \\
\hline 
$(\bar{x}_0, \bar{x}_1) = D_\theta\left(x_{\alpha_t}, {\alpha_t}\right)$ & 
$\bar{x}_0 = D_\theta\left(x_{\alpha_t}, {\alpha_t}\right)$ & 
$\bar{x}_1 = D_\theta\left(x_{\alpha_t}, {\alpha_t}\right)$ & 
$\bar{x}_1-\bar{x}_0 = D_\theta\left(x_{\alpha_t}, {\alpha_t}\right)$ 
\\
\hline
$x_{\alpha_{t+1}} = $ &
$x_{\alpha_{t+1}} = \bar{x}_{0}~+ $ &
$x_{\alpha_{t+1}} = \bar{x}_{1}~+ $ &
$x_{\alpha_{t+1}} = x_{\alpha_{t}} + $
\\
$(1-\alpha_{t+1})\,\bar{x}_0 + \alpha_{t+1} \bar{x}_1 $ &
$\frac{\alpha_{t+1}}{\alpha_{t}}\,(x_{\alpha_{t}} - \bar{x}_0)$ &
$\frac{(1-\alpha_{t+1})}{(1-\alpha_{t})}\,(x_{\alpha_{t}} - \bar{x}_1)$ &
$(\alpha_{t+1} - \alpha_{t}) \, (\bar{x}_1-\bar{x}_0)$
\\
\hline
{\color{myred}unstable} &
{\color{myred}unstable when $\alpha_t \rightarrow 0$} &
{\color{myred}unstable when $\alpha_t \rightarrow 1$} &
{\color{mygreen}stable}
\\
\hline
\end{tabular}
\end{tabular}
}
\vspace{1mm}
\caption{\label{tab:equivalent_formulations} 
\textbf{Variant formulations of iterative $\alpha$-(de)blending} (equivalent in theory but not in practice).}
\vspace{-4mm}
\end{table}

\subsection{Training and Sampling}
\label{sec:training_sampling}

Following variant (d) of Table~\ref{tab:equivalent_formulations}, we train the neural network $D_\theta$ to predict the average difference vector between the posterior samples. Our learning objective is defined by 
\begin{align}
\displaystyle
\min_{\theta}~~\mathop{\mathbb{E}}_{\alpha, x_\alpha}
\left[\left\|D_\theta\left(x_\alpha, \alpha\right)  
- 
\mathop{\mathbb{E}}_{\left(x_0, x_1\right)_{| (x_{\alpha}, \alpha)}}\left[x_1 - x_0\right]\right\|^2\right].
\end{align}
Note that minimizing the $l^2$ norm of the average of a distribution is equivalent to minimizing the $l^2$ norm of all of the samples of the distribution. 
We obtain the equivalent objective
\begin{align}
\displaystyle
\min_{\theta}~~\mathop{\mathbb{E}}_{\alpha, x_\alpha, \left(x_0, x_1\right)_{| (x_{\alpha}, \alpha)}}
\left[\left\|D_\theta\left(x_\alpha, \alpha\right)  - \left(x_1 - x_0\right)\right\|^2\right].
\end{align}
Finally, as explained in Section~\ref{sec:mapping_preliminaries}, sampling $x_\alpha \sim p_\alpha$ first and then $\left(x_0, x_1\right)_{| (x_{\alpha}, \alpha)}$ is equivalent to sampling $\left(x_0,x_1\right) \sim \left(p_0,p_1\right)$ and blending them to obtain $x_\alpha \sim p_\alpha$.
With this, we obtain our final learning objective
\begin{align}
\displaystyle
\min_{\theta}~~\mathop{\mathbb{E}}_{\alpha, x_0, x_1}
\left[\left\|D_\theta\left((1-\alpha) \, x_0 + \alpha \, x_1, \alpha\right)  - \left(x_1-x_0\right)\right\|^2\right],
\label{eq:final_objective}
\end{align}
which we use to optimize $\theta$ in Algorithm~\ref{algo:training}.
In Algorithm~\ref{algo:sampling}, we iteratively map samples $x_0 \sim p_0$ to samples $x_1 \sim p_1$ in the same way as in Algorithm~\ref{algo:iterative_deterministic_sampling}, where we use the neural network $D_\theta$ to obtain the average posterior difference.

\begin{algorithm}
\caption{Training}\label{algo:training}
\begin{algorithmic}
\Require $x_0 \sim p_0$, $x_1 \sim p_1$, $\alpha \sim \mathcal{U}_{[0,1]}$
\State $x_\alpha = (1-\alpha) \, x_0 + \alpha \, x_1$
\State $l = \|D_\theta\left(x_\alpha, \alpha\right) - (x_1-x_0)\|^2$
\State backprop from $l$ and update $\theta$
\end{algorithmic}
\end{algorithm}

\vspace{-2mm}

\begin{algorithm}
\caption{Sampling}\label{algo:sampling}
\begin{algorithmic}
\Require $x_0 \sim p_0$, $T$, $\alpha_t := \frac{t}{T}$
\For{$t=0,..,T-1$}
\State $x_{\alpha_{t+1}} = x_{\alpha_t} + \left(\alpha_{t+1}-\alpha_{t}\right) \, D_\theta\left(x_{\alpha_t}, \alpha_t\right)$
\EndFor
\end{algorithmic}
\end{algorithm}

\section{Experiments with Analytic Densities}
\label{sec:experiments}

\paragraph{Experiments with 1D densities.}

In Figure~\ref{fig:l2_vs_l1}, we experiment with analytic 1D densities, where the expectation $\bar{x}_1-\bar{x}_0$ can be computed analytically rather than being learnt by a neural network $D_\theta$.
The experiment confirms that the analytic version matches the reference and that the neural network trained with the $l_2$ norm approximates the same mapping. 
We also tested training the neural network with the $l_1$ norm, which makes the neural network approximate the median of $x_1-x_0$ rather than its average.
The resulting mapping does not match the reference. 
This confirms that learning the average via $l_2$~training is a key component of our model, as explained in Section~\ref{sec:training_sampling}.

\paragraph{Experiments with 2D densities.}

Figure~\ref{fig:results_2d_intermediate} shows that the intermediate blended densities $p_\alpha$ computed by our mapping match the reference blended densities. 
Figure~\ref{fig:results_2d_mapping} shows how our algorithm maps the samples of $p_0$ to samples of $p_1$.
These results demonstrate that IADB computes valid mappings between arbitrary densities.

\section{Experiments with Image Densities}
\label{sec:nongaussian}

\subsection{Gaussian sampling: comparison against DDIM}
\label{sec:ddim}

The state-of-the-art deterministic denoising diffusion model is DDIM \cite{song2021ddim}.
The derivation of DDIM is based on the assumption that the first density is Gaussian.
In this section, we show that if $p_0$ is Gaussian then IADB produces the same mapping as DDIM but with a different parameterization (a given $x_0$ is mapped to the same $x_1$ but the trajectory is different).

\paragraph{\normalfont\textbf{Proposition.}}

If $p_0$ is a Gaussian density, IADB and DDIM define the same deterministic mapping.

\paragraph{\normalfont\textbf{Proof.}}

In Appendix~C of our supplemental, we show that, with a simple change of parameterization, the update rule of DDIM corresponds to variant (b) in Table~\ref{tab:equivalent_formulations}.

\paragraph{IADB consistently outperforms DDIM in FID scores.}

In Figure~\ref{fig:res_cmp_ours_ddim}, we experiment under the same conditions (architecture, training time, 1st-order solver, uniform schedule, Gaussian $p_0$) and measure the FID score~\citep{heusel2017} for a varying number of sampling steps on 3 image datasets: LSUN Bedrooms (64x64), CelebA (64 x 64) and AFHQ Cats(128x128).
We use a U-Net architecture from the HuggingFace Diffusers library\footnote{https://github.com/huggingface/diffusers}.
The model has 6 downsampling blocks, 6 upsampling blocks, and a self-attention middle block.
We trained the model with the AdamW optimizer (Learning rate=0.0001, Weight decay=0.01, betas=(0.9,0.999)) and we set the batch size to 64 for Celeba, 8 for AFHQ Cats, 64 for bedrooms, and 128 for Cifar.
All models are trained for 120 hours of training (approx. 800k steps on the CelebA dataset) on a single NVIDIA Titan RTX. 
We observe a consistently better performance with IADB compared to DDIM.

\paragraph{Discussion.}

The improved performance of IADB compared to DDIM is due to multiple factors. 
The formulation generally used in DDIM corresponds to variant (b) presented in Table~\ref{tab:equivalent_formulations}: they train a denoiser to predict the Gaussian noise present in the noisy image samples, i.e., their model learns to predict $\bar{x}_0$. 
However, we explain that this variant makes the sampling less stable because of the division near 0.
As a matter of fact, in their implementation, the sampler starts at some $\epsilon>0$ precisely to avoid dividing by 0.
Our variant (d) does not suffer from this problem.
Another factor is that the learning objective defined by variant (d) provides a better optimization landscape than variant (b).
For instance, the effort to learn $\bar{x}_0$ in variant (b) is imbalanced over $\alpha$ because the $l^2$ norm is small near $\alpha=0$ and large near $\alpha=1$. 
In contrast, the effort to learn $\bar{x}_1-\bar{x}_0$ in variant (d) is more balanced over $\alpha$.

\subsection{Non-Gaussian sampling}

In contrast to DDIM, IADB does not make the assumption that $p_0$ is Gaussian.
Indeed, IADB is theoretically proven to produce a correct sampling of $p_1$ for \textit{any} $p_0$ (as long as they are Riemann integrable and of finite variance).
In this section, we experiment with how IADB behaves with non-Gaussian densities for $p_0$.

\paragraph{Impact on sampling quality.}

In the experiment of Figure~\ref{fig:res_nongaussian}, we use IADB to sample face images ($p_1$) using different densities for $p_0$.
We observe that IADB does indeed generate faces regardless of the choice of $p_0$.
However, while we observe a similar sampling quality for analytic $p_0$ (Gaussian, uniform and bi-Gaussian noises), we see a significant drop in quality when $p_0$ is not an analytic primitive but a real-image dataset such as the pebble textures.
This observation does not invalidate the theory: IADB effectively defines a valid mapping between the pebble density $p_0$ and the face density $p_1$, but the quality achieved is lower in practice.
We conjecture that this is because the learning task is more difficult.
Indeed, the analytic noise primitives provide a simple and smooth landscape for $p_0$ such that the learning capacity of the neural network can be entirely spent on learning the $p_1$ manifold.
In contrast, when $p_0$ is also a complex image manifold, the learning capacity of the network is spent on both $p_0$ and $p_1$.
This might explain the lower quality when generating samples from $p_1$. 

\paragraph{Correct but unfaithful mappings.}

Figure~\ref{fig:restoration} shows an experiment where we use IADB to sample color face images ($p_1$) using grayscale face images ($p_0$).
We observe that IADB successfully accomplishes this task: it effectively maps grayscale faces to color faces.
Unfortunately, the output color images are not colorizations of the input grayscale images, which rules out some user applications.
However, this is not the promise of the theory. 
Indeed, the theory predicts that the mapping produces a valid sampling of $p_1$ using $p_0$ (which is the case in the experiment) but not that the mapping will be what a user expects (which is not the case in the experiment). 

\paragraph{Faithful mappings with conditional IADB} 

Previous works show that conditioning the diffusion process seems to be necessary for achieving faithful image-to-image translations.
For instance, adding an energy guide during the ODE integration~\citep{zhao2022egsde} by progressively injecting features~\citep{meng2021sdedit} or by sampling conditional densities~\citet{saharia2021superres,saharia2022palette}.
In Figure~\ref{fig:res_conditional}, we experiment with the latter using Gaussian noise for $x_0$, clean images for $x_1$, and a corrupted version of $x_1$ for the condition $c$ passed as an additional argument to the neural network: $D_\theta(c, x_\alpha, \alpha) = \bar{x}_1-\bar{x}_0$.
The experiment shows that IADB can be successfully used to obtain faithful image-to-image translations with additional conditioning.

\pagebreak

\section{Discussion}
\label{sec:discussion}

\paragraph{Improved sampler.}

We experimented with IADB in its vanilla setting with a uniform blending schedule and a first-order ODE solver. 
It readily benefits from orthogonal improvements brought to denoising diffusion, such as better blending schedules and higher-order ODE solvers~\citep{karras2022}.
For instance, Algorithm~\ref{algo:sampling_improved} provides an improved version of Algorithm~\ref{algo:sampling} with a 2nd-order Runge-Kutta integration and a cosine schedule. 

\begin{algorithm}
\caption{Sampling (2nd-order Runge-Kutta, cosine schedule)}\label{algo:sampling_improved}
\begin{algorithmic}
\Require $x_0 \sim p_0$, $T$, $\alpha_t := 1 - \cos\left(\frac{t}{T} \, \frac{\pi}{2}\right)$
\For{$t=0,..,T-1$}
\State $x_{\alpha_{t+\frac{1}{2}}} = x_{\alpha_t} + \left(\alpha_{t+\frac{1}{2}}-\alpha_t\right) \, D_\theta\left(x_{\alpha_t}, \alpha_t\right)$
\State $x_{\alpha_{t+1}} = x_{\alpha_t} + \left(\alpha_{t+1}-\alpha_{t}\right) \, D_\theta\left(x_{\alpha_{t+\frac{1}{2}}}, \alpha_{t+\frac{1}{2}}\right)$
\EndFor
\end{algorithmic}
\end{algorithm}

\paragraph{Stochastic Differential Equations (SDEs).}

The random sequence computed by the stochastic version of IADB presented in Algorithm~\ref{algo:iterative_stochastic_sampling} is a Markov chain.
This algorithm might therefore appear reminiscent of stochastic diffusion models~\cite{song2021sde} based on SDEs.
However, IADB is not related to an SDE.
Indeed, SDEs model stochastic behaviors at the infinitesimal scale while our mapping is stochastic for discrete steps and becomes a deterministic ODE in the infinitesimal limit.

\paragraph{Non-Gaussian denoising diffusion.}

Some previous works have focused on replacing Gaussian noise with other noise distributions, such as the generalised normal (exponential power) distribution~\cite{deasy2021heavytailed} or the Gamma distribution~\cite{nachmani2021gamma}.
Our more general derivation works with any finite-variance density rather than specific noise alternatives.
Peluchetti~\shortcite{peluchetti2022nondenoising} proposes a more general SDE framework. 
Our ODE can be derived from his SDE by nullifying the stochastic component and following its aggregation method. 
In this respect, our ODE is not entirely new.
However, our derivation is new and incomparably simpler, which is the main point of this paper.

\section{Conclusion}
\label{sec:conclusion}

The objective of this work was to find a simple and intuitive way to approach deterministic denoising diffusion. 
Using only simple sampling concepts, we derived Iterative $\alpha$-(De)Blending (IADB), a deterministic diffusion model based on a sampling interpretation of blending and deblending.
We have seen that our model defines exactly the same mapping as DDIM~\cite{song2021ddim}, the state-of-the-art competitor in deterministic denoising diffusion.
This yields a positive answer to the question asked in the introduction \textit{``Is there a simpler approach to deterministic diffusion?''}.
Indeed, it is possible to derive the same result without leveraging any knowledge about Langevin dynamics, score matching, SDEs, etc. 
Getting there was the whole point of this paper. 
Furthermore, our simpler IADB derivation provides both practical and theoretical gains. 
It has led to a more numerically stable formulation that produces better FID scores than DDIM and has revealed that DDIM's Gaussian assumption is theoretically unnecessary.

\bibliographystyle{ACM-Reference-Format}
\bibliography{file_BIBLIO}
\clearpage

\appendix

\clearpage

\begin{figure}[!h]
    \center
    \begin{tikzpicture}
        \begin{scope}[shift={(0.0, 0\linewidth)}]
        \node[inner sep=0pt] (ref) {\includegraphics[width=0.22\linewidth, frame=0.1pt, rotate=90]{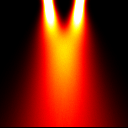}};
        \begin{scope}[rotate=90]
            \begin{axis}[
                name=upperPlot,
                width=0.4\linewidth,
                height=2.5cm,
                samples=512,
                xticklabels=\empty,
                yticklabels=\empty,
                grid=major,
                xmin = -2,
                xmax = 2,
                anchor=south,
                at = {(ref.west)},
                yshift=2pt,
            ]
                \addplot[color=red, thick]{0.5*(exp(-0.5*(x+0.5)^2/0.01) + exp(-0.5*(x-0.5)^2/0.01))};
            \end{axis}
        \end{scope}
        \begin{scope}[rotate=90]
            \begin{axis}[
                name=lowerPlot,
                width=0.4\linewidth,
                height=2.5cm,
                samples=128,
                xticklabels=\empty,
                yticklabels=\empty,
                grid=major,
                xmin = -2,
                xmax = 2,
                anchor=north,
                at = {(ref.east)},
                yshift=-2pt,
            ]
                \addplot[color=blue, thick]{exp(-0.5*(x)^2)};
            \end{axis}
        \end{scope}

        \begin{scope}[xshift=-3.2cm, yshift=-3.0cm]
        \node[inner sep=0pt, rotate=90, xshift=1pt] (L2) {\includegraphics[width=0.22\linewidth, frame=0.1pt]{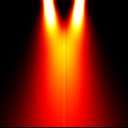}};
        \node[inner sep=0pt, right=5pt of L2.south east, anchor=north east, rotate=90] (neural_L2)
        {\includegraphics[width=0.22\linewidth, frame=0.1pt]{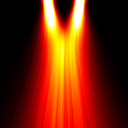}};
        
        \node[inner sep=0pt, right=10pt of neural_L2.south east, anchor=north east, rotate=90] (L1) 
        {\includegraphics[width=0.22\linewidth, frame=0.1pt]{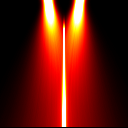}}; 
        \node[inner sep=0pt, right=5pt of L1.south east, anchor=north east, rotate=90] (neural_L1) {\includegraphics[width=0.22\linewidth, frame=0.1pt]{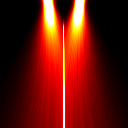}};

        \draw[|-|] ($(L2.north west)+(0.05,-0.1)$) --  ($(neural_L1.south west)+(-0.05,-0.1)$) node [midway, below] { {\footnotesize IADB with analytic expressions or neural networks (nn) for $D_\theta$} };
        
        \node[above=0pt of L2.east,anchor=south] {\footnotesize analytic average};
        \node[above=0pt of neural_L2.east,anchor=south] {\footnotesize $D_\theta$ ($l_2$ training)};
        \node[above=0pt of L1.east,anchor=south] {\footnotesize analytic median};
        \node[above=0pt of neural_L1.east,anchor=south] {\footnotesize $D_\theta$ ($l_1$ training)};
        \end{scope}

        \draw[->] ($(ref.south west)+(0,-0.1)$) -- ($(ref.south east)+(0,-0.1)$) node[midway, below, shift={(0,-0.01)}] {\footnotesize increasing $\alpha$};
        \node[above=2pt of ref, anchor=south] { {\footnotesize reference } };
    \end{scope}
        \begin{scope}[shift={(0.0, -0.75\linewidth)}]
        \node[inner sep=0pt] (ref) {\includegraphics[width=0.22\linewidth, frame=0.1pt, rotate=90]{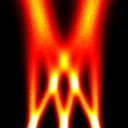}};
        \begin{scope}[rotate=90]
            \begin{axis}[
                name=upperPlot,
                width=0.4\linewidth,
                height=2.5cm,
                samples=512,
                xticklabels=\empty,
                yticklabels=\empty,
                grid=major,
                xmin = -2,
                xmax = 2,
                anchor=south,
                at = {(ref.west)},
                yshift=2pt,
            ]
                \addplot[color=red, thick]{0.5*(exp(-0.5*(x+0.9)^2/0.3^2) + exp(-0.5*(x-0.9)^2/0.3^2))};
            \end{axis}
        \end{scope}
        \begin{scope}[rotate=90]
            \begin{axis}[
                name=lowerPlot,
                width=0.4\linewidth,
                height=2.5cm,
                samples=128,
                xticklabels=\empty,
                yticklabels=\empty,
                grid=major,
                xmin = -2,
                xmax = 2,
                anchor=north,
                at = {(ref.east)},
                yshift=-2pt,
            ]
                \addplot[color=blue, thick, samples=512]{0.5*(exp(-0.5*(x+0.0)^2/0.1^2) + exp(-0.5*(x-1.0)^2/0.1^2) + exp(-0.5*(x+1.0)^2/0.1^2))};
            \end{axis}
        \end{scope}

        \begin{scope}[xshift=-3.2cm, yshift=-3.0cm]
        \node[inner sep=0pt, rotate=90, xshift=1pt] (L2) {\includegraphics[width=0.22\linewidth, frame=0.1pt]{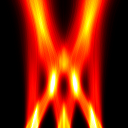}};
        \node[inner sep=0pt, right=5pt of L2.south east, anchor=north east, rotate=90] (neural_L2)
        {\includegraphics[width=0.22\linewidth, frame=0.1pt]{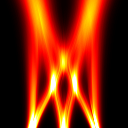}};
        
        \node[inner sep=0pt, right=10pt of neural_L2.south east, anchor=north east, rotate=90] (L1) 
        {\includegraphics[width=0.22\linewidth, frame=0.1pt]{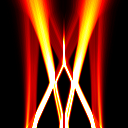}}; 
        \node[inner sep=0pt, right=5pt of L1.south east, anchor=north east, rotate=90] (neural_L1) {\includegraphics[width=0.22\linewidth, frame=0.1pt]{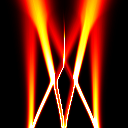}};

        \draw[|-|] ($(L2.north west)+(0.05,-0.1)$) --  ($(neural_L1.south west)+(-0.05,-0.1)$) node [midway, below] { {\footnotesize IADB with analytic expressions or neural networks (nn) for $D_\theta$} };
        
        \node[above=0pt of L2.east,anchor=south] {\footnotesize analytic average};
        \node[above=0pt of neural_L2.east,anchor=south] {\footnotesize $D_\theta$ ($l_2$ training)};
        \node[above=0pt of L1.east,anchor=south] {\footnotesize analytic median};
        \node[above=0pt of neural_L1.east,anchor=south] {\footnotesize $D_\theta$ ($l_1$ training)};
        \end{scope}

        \draw[->] ($(ref.south west)+(0,-0.1)$) -- ($(ref.south east)+(0,-0.1)$) node[midway, below, shift={(0,-0.01)}] {\footnotesize increasing $\alpha$};
        \node[above=2pt of ref, anchor=south] { {\footnotesize reference } };
    \end{scope}
    \end{tikzpicture}
    \vspace{-3mm}
    \caption{ \label{fig:l2_vs_l1}
    \textbf{Experiments with 1D densities.}
        (top) We blend a bi-Normal distribution with modes $\mu_{1|2} = \{-0.5, 0.5\}$ and $\sigma_{1|2} = 0.1$ (in red) with a Normal distribution of unit variance (in blue).
        (bottom) We blend a bi-Normal distribution with modes $\mu_{1|2} = \{-0.9, 0.9\}$ and $\sigma_{1|2} = 0.3$ (in red) to a tri-Normal distribution with $\mu_{1|2|3} = \{-1, 0, 1\}$ and $\sigma_{1|2|3} = 0.1$  (in blue). 
        The reference shows analytically convolved densities $p_\alpha$.
        The other densities are the histograms of the samples $x_\alpha$ computed in Algorithm~\ref{algo:sampling} using either analytic expressions or neural networks for $D_\theta$. 
        The neural network is an MLP with 5 hidden layers of 64 filters trained with the Adam optimizer with a learning rate of $10^{-5}$ for 10k iterations.
    }
\end{figure}
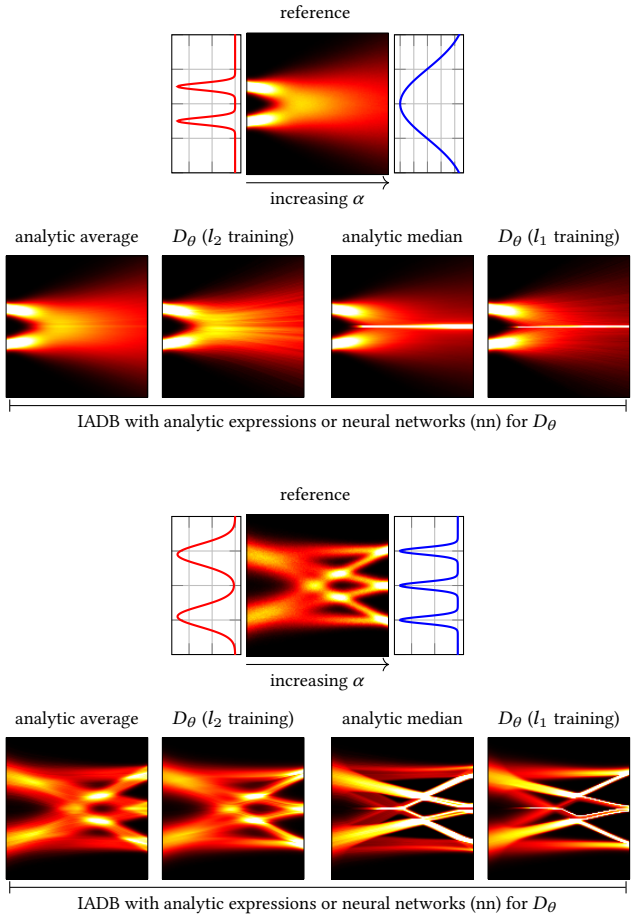

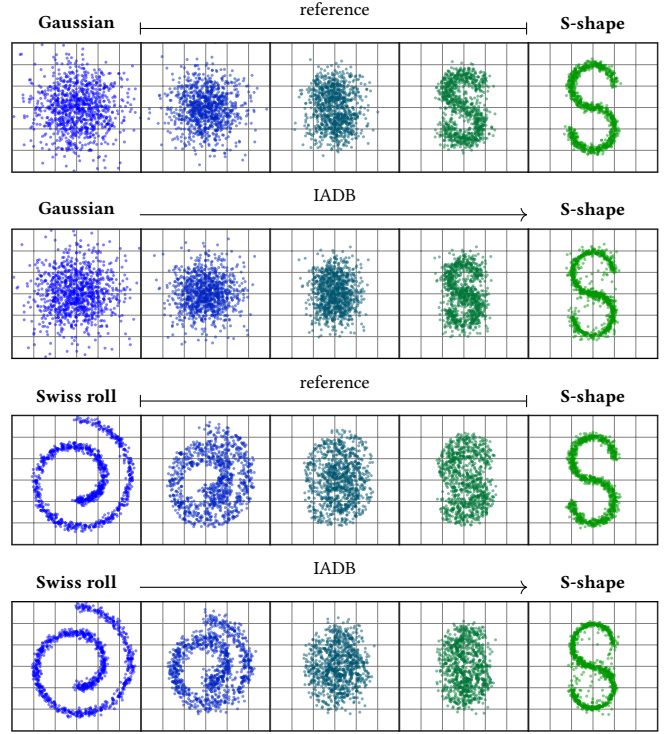
\begin{figure}[!h]
    \hspace{-4pt}
    \begin{tikzpicture}
        \begin{scope}[shift={(0.0, 0.0)}]
            \begin{scope}
                \node[inner sep=0pt,                 ] (A0) { \includegraphics[width=0.2\linewidth, frame]{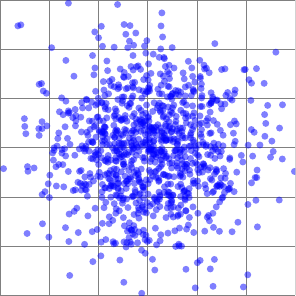} };
                \node[inner sep=0pt, right=-1pt of A0] (A1) { \includegraphics[width=0.2\linewidth, frame]{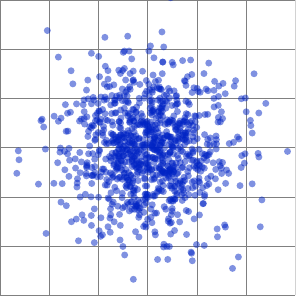} };
                \node[inner sep=0pt, right=-1pt of A1] (A2) { \includegraphics[width=0.2\linewidth, frame]{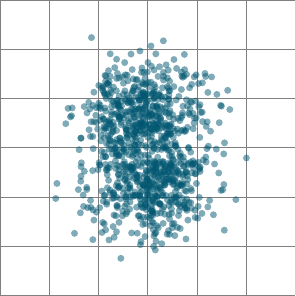} };
                \node[inner sep=0pt, right=-1pt of A2] (A3) { \includegraphics[width=0.2\linewidth, frame]{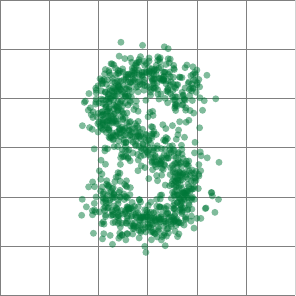} };
                \node[inner sep=0pt, right=-1pt of A3] (A4) { \includegraphics[width=0.2\linewidth, frame]{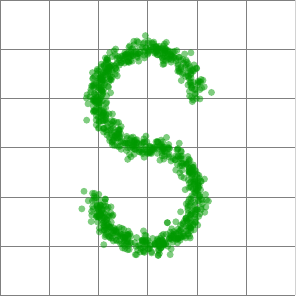} };
                
                \node[above=13pt of A0, anchor=north] { {\footnotesize \textbf{Gaussian}} };
                \node[above=18pt of A2, anchor=north] { {\footnotesize reference} };
                \node[above=13pt of A4, anchor=north] { {\footnotesize \textbf{S-shape}} };
                \draw[|-|] ($(A1.north west) + (0, 5pt)$) -- ($(A4.north west) + (0, 5pt)$);
            \end{scope}
        \end{scope}
        \begin{scope}[shift={(0.0, -0.29\linewidth)}]
            \begin{scope}
                \node[inner sep=0pt,                 ] (A0) { \includegraphics[width=0.2\linewidth, frame]{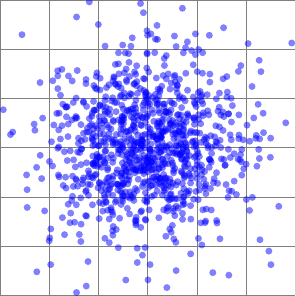} };
                \node[inner sep=0pt, right=-1pt of A0] (A1) { \includegraphics[width=0.2\linewidth, frame]{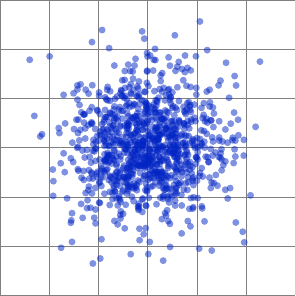} };
                \node[inner sep=0pt, right=-1pt of A1] (A2) { \includegraphics[width=0.2\linewidth, frame]{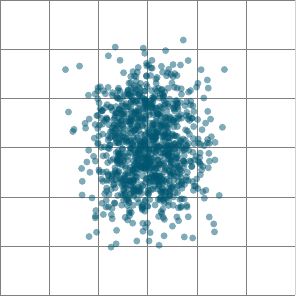} };
                \node[inner sep=0pt, right=-1pt of A2] (A3) { \includegraphics[width=0.2\linewidth, frame]{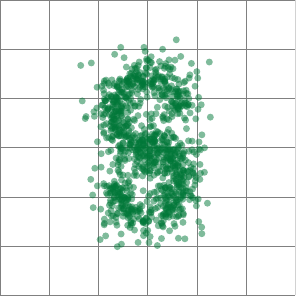} };
                \node[inner sep=0pt, right=-1pt of A3] (A4) { \includegraphics[width=0.2\linewidth, frame]{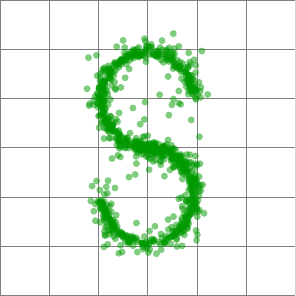} };
                
                \node[above=13pt of A0, anchor=north] { {\footnotesize \textbf{Gaussian}} };
                \node[above=18pt of A2, anchor=north] { {\footnotesize IADB} };
                \node[above=13pt of A4, anchor=north] { {\footnotesize \textbf{S-shape}} };
                \draw[->] ($(A1.north west) + (0, 5pt)$) -- ($(A4.north west) + (0, 5pt)$);
            \end{scope}
        \end{scope}        
        \begin{scope}[shift={(0.0\linewidth, -0.58\linewidth)}]
            \begin{scope}
                \node[inner sep=0pt,                 ] (A0) { \includegraphics[width=0.2\linewidth, frame]{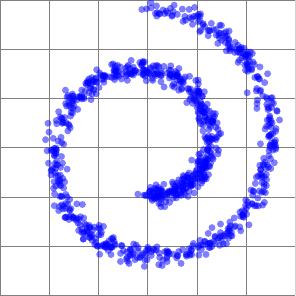} };
                \node[inner sep=0pt, right=-1pt of A0] (A1) { \includegraphics[width=0.2\linewidth, frame]{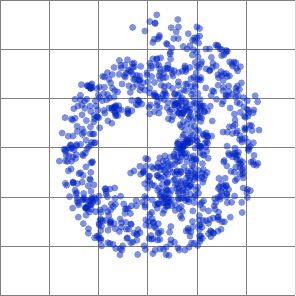} };
                \node[inner sep=0pt, right=-1pt of A1] (A2) { \includegraphics[width=0.2\linewidth, frame]{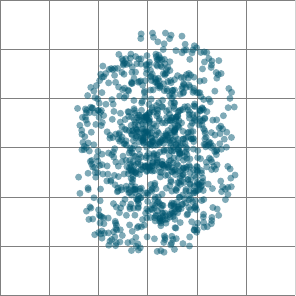} };
                \node[inner sep=0pt, right=-1pt of A2] (A3) { \includegraphics[width=0.2\linewidth, frame]{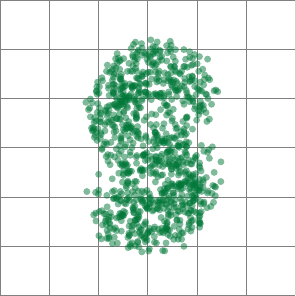} };
                \node[inner sep=0pt, right=-1pt of A3] (A4) { \includegraphics[width=0.2\linewidth, frame]{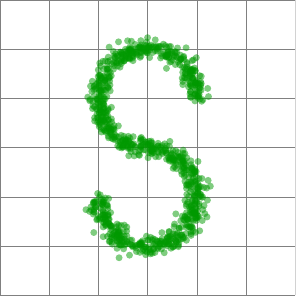} };
                
                \node[above=13pt of A0, anchor=north] { {\footnotesize \textbf{Swiss roll}} };
                \node[above=18pt of A2, anchor=north] { {\footnotesize reference} };
                \node[above=13pt of A4, anchor=north] { {\footnotesize \textbf{S-shape}} };
                \draw[|-|] ($(A1.north west) + (0, 5pt)$) -- ($(A4.north west) + (0, 5pt)$);
            \end{scope}
        \end{scope}
        \begin{scope}[shift={(0.0\linewidth, -0.87\linewidth)}]
            \begin{scope}
                \node[inner sep=0pt,                 ] (A0) { \includegraphics[width=0.2\linewidth, frame]{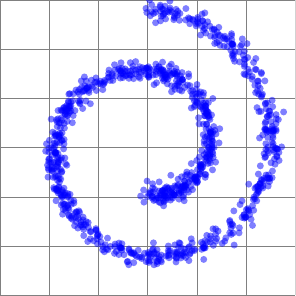} };
                \node[inner sep=0pt, right=-1pt of A0] (A1) { \includegraphics[width=0.2\linewidth, frame]{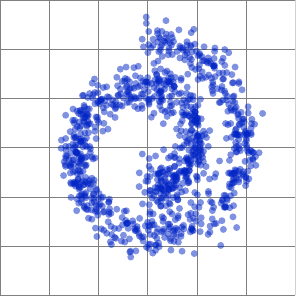} };
                \node[inner sep=0pt, right=-1pt of A1] (A2) { \includegraphics[width=0.2\linewidth, frame]{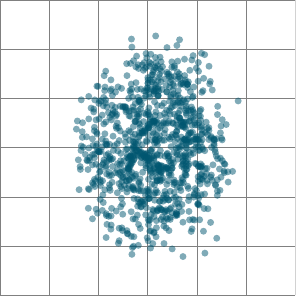} };
                \node[inner sep=0pt, right=-1pt of A2] (A3) { \includegraphics[width=0.2\linewidth, frame]{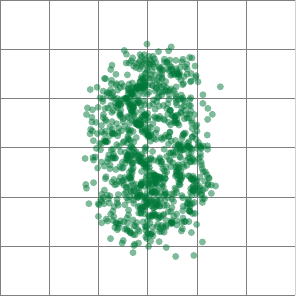} };
                \node[inner sep=0pt, right=-1pt of A3] (A4) { \includegraphics[width=0.2\linewidth, frame]{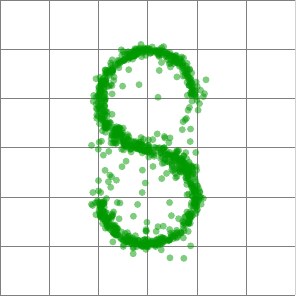} };
                
                \node[above=13pt of A0, anchor=north] { {\footnotesize \textbf{Swiss roll}} };
                \node[above=18pt of A2, anchor=north] { {\footnotesize IADB} };
                \node[above=13pt of A4, anchor=north] { {\footnotesize \textbf{S-shape}} };
                \draw[->] ($(A1.north west) + (0, 5pt)$) -- ($(A4.north west) + (0, 5pt)$);
            \end{scope}
        \end{scope}
    \end{tikzpicture}
    \vspace{-6mm}
    \caption{\label{fig:results_2d_intermediate}
    \textbf{Experiments with 2D densities: intermediate distributions.}
We show samples of the intermediate densities $p_\alpha$. References are computed by $\alpha$-blending random samples from $p_0$ and $p_1$. IADB intermediate distributions are computed with Algorithm~\ref{algo:sampling} using an MLP with 5 hidden layers of 64 filters for $D_\theta$ trained with the Adam optimizer with a learning rate of $10^{-5}$ for 10k iterations.
    }
    \vspace{6mm}
\end{figure}

\begin{figure}[!h]
    \hspace{-12pt}
    {\small
    \begin{tikzpicture}
        \node[inner sep=0] (A0) { \includegraphics[width=0.32\linewidth, frame]{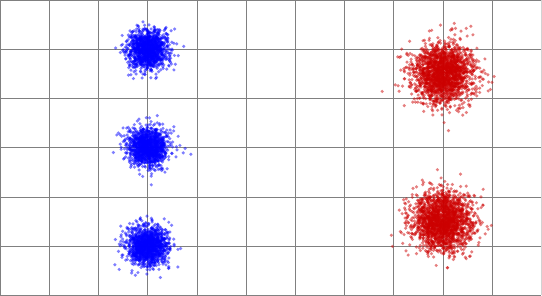} };
        \node[inner sep=0, right=2pt of A0] (A1) { \includegraphics[width=0.32\linewidth, frame]{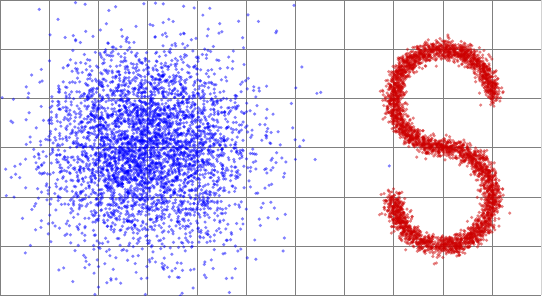} };
        \node[inner sep=0, right=2pt of A1] (A2) { \includegraphics[width=0.32\linewidth, frame]{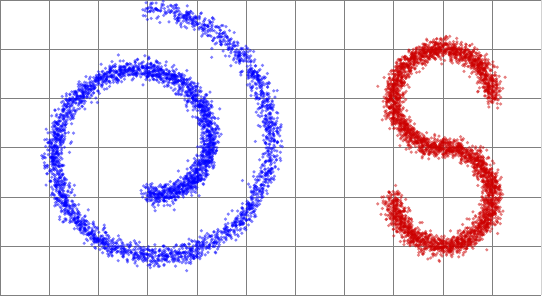} };
        \node[inner sep=0, left=4pt of A0, anchor=south, rotate=90] {$p_0$ and $p_1$};
        
        \node[inner sep=0, below=2pt of A0] (B0) { \includegraphics[width=0.32\linewidth, frame]{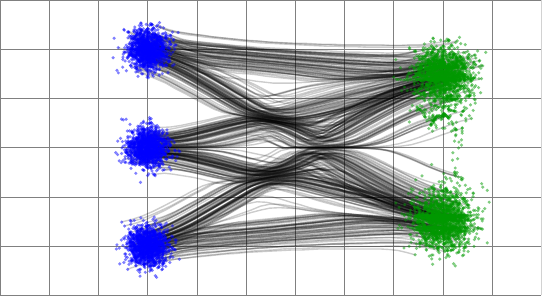} };
        \node[inner sep=0, right=2pt of B0] (B1) { \includegraphics[width=0.32\linewidth, frame]{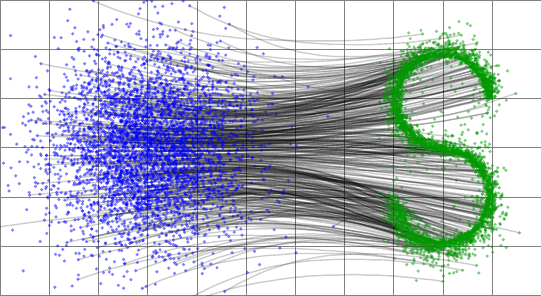} };
        \node[inner sep=0, right=2pt of B1] (B2) { \includegraphics[width=0.32\linewidth, frame]{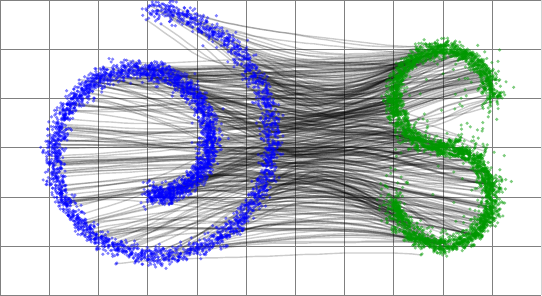} };
        \node[inner sep=0, left=3pt of B0, anchor=south, rotate=90] {IADB};
    
        \node[inner sep=0, above=10pt of A0, anchor=north] {3-Normal to 2-Normal};    
        \node[inner sep=0, above=10pt of A1, anchor=north] {Normal to S-curve};    
        \node[inner sep=0, above=10pt of A2, anchor=north] {Roll to S-curve};    
    \end{tikzpicture}
    }
    \vspace{-6mm}
\caption{\label{fig:results_2d_mapping}
\textbf{Experiments with 2D densities: mappings.}
We show samples of the reference densities $p_0$ and $p_1$ and the mapping computed by Algorithm~\ref{algo:sampling} using an MLP with 5 hidden layers of 64 filters for $D_\theta$ trained with the Adam optimizer with a learning rate of $10^{-5}$ for 10k iterations.
The final samples computed by the algorithm (green) match the reference samples $x_1$ (red).
}
\end{figure}
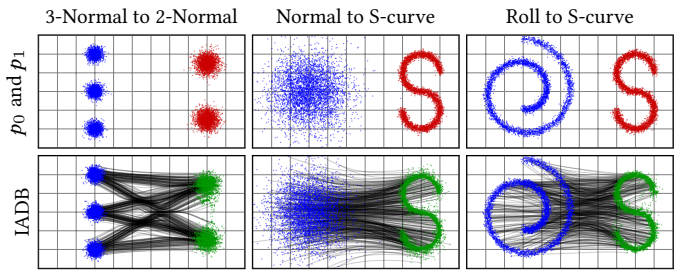

\clearpage
\input{figures/fig_RESULTS_CMP_DDIM.tex}

\begin{figure}[!h]
    \hspace{-5pt}
    \begin{tikzpicture}
        \begin{scope}[shift={(0.0, +0.27\linewidth)}]
            \begin{scope}
                \node[inner sep=0pt,                 ] (A0) { \includegraphics[width=0.2\linewidth, frame]{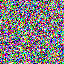} };
                \node[inner sep=0pt, right=-1pt of A0] (A1) { \includegraphics[width=0.2\linewidth, frame]{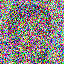} };
                \node[inner sep=0pt, right=-1pt of A1] (A2) { \includegraphics[width=0.2\linewidth, frame]{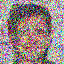} };
                \node[inner sep=0pt, right=-1pt of A2] (A3) { \includegraphics[width=0.2\linewidth, frame]{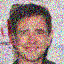} };
                \node[inner sep=0pt, right=-1pt of A3] (A4) { \includegraphics[width=0.2\linewidth, frame]{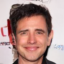} };
                \node[above=13pt of A0, anchor=north] { {\footnotesize \textbf{Gaussian}} };
                \node[above=13pt of A4, anchor=north] { {\footnotesize \textbf{CelebA}} };
                \draw[->] ($(A1.north west) + (0, 5pt)$) -- ($(A4.north west) + (0, 5pt)$);
            \end{scope}
        \end{scope}
        \begin{scope}[shift={(0.0, +0.0\linewidth)}]
            \begin{scope}
                \node[inner sep=0pt,                 ] (A0) { \includegraphics[width=0.2\linewidth, frame]{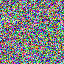} };
                \node[inner sep=0pt, right=-1pt of A0] (A1) { \includegraphics[width=0.2\linewidth, frame]{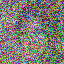} };
                \node[inner sep=0pt, right=-1pt of A1] (A2) { \includegraphics[width=0.2\linewidth, frame]{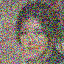} };
                \node[inner sep=0pt, right=-1pt of A2] (A3) { \includegraphics[width=0.2\linewidth, frame]{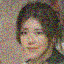} };
                \node[inner sep=0pt, right=-1pt of A3] (A4) { \includegraphics[width=0.2\linewidth, frame]{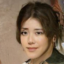} };
                \node[above=13pt of A0, anchor=north] { {\footnotesize \textbf{Uniform}} };
                \node[above=13pt of A4, anchor=north] { {\footnotesize \textbf{CelebA}} };
                \draw[->] ($(A1.north west) + (0, 5pt)$) -- ($(A4.north west) + (0, 5pt)$);
            \end{scope}
        \end{scope}
        \begin{scope}[shift={(0.0, -0.27\linewidth)}]
            \begin{scope}
                \node[inner sep=0pt,                 ] (A0) { \includegraphics[width=0.2\linewidth, frame]{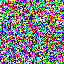} };
                \node[inner sep=0pt, right=-1pt of A0] (A1) { \includegraphics[width=0.2\linewidth, frame]{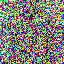} };
                \node[inner sep=0pt, right=-1pt of A1] (A2) { \includegraphics[width=0.2\linewidth, frame]{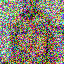} };
                \node[inner sep=0pt, right=-1pt of A2] (A3) { \includegraphics[width=0.2\linewidth, frame]{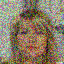} };
                \node[inner sep=0pt, right=-1pt of A3] (A4) { \includegraphics[width=0.2\linewidth, frame]{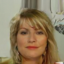} };
                \node[above=13pt of A0, anchor=north] { {\footnotesize \textbf{Bi-Gaussian}} };
                \node[above=13pt of A4, anchor=north] { {\footnotesize \textbf{CelebA}} };
                \draw[->] ($(A1.north west) + (0, 5pt)$) -- ($(A4.north west) + (0, 5pt)$);
            \end{scope}
        \end{scope}
        \begin{scope}[shift={(0.0, -0.54\linewidth)}]
            \begin{scope}
                \node[inner sep=0pt,                 ] (A0) { \includegraphics[width=0.2\linewidth, frame]{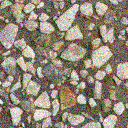} };
                \node[inner sep=0pt, right=-1pt of A0] (A1) { \includegraphics[width=0.2\linewidth, frame]{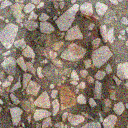} };
                \node[inner sep=0pt, right=-1pt of A1] (A2) { \includegraphics[width=0.2\linewidth, frame]{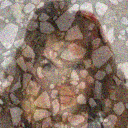} };
                \node[inner sep=0pt, right=-1pt of A2] (A3) { \includegraphics[width=0.2\linewidth, frame]{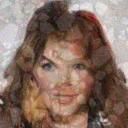} };
                \node[inner sep=0pt, right=-1pt of A3] (A4) { \includegraphics[width=0.2\linewidth, frame]{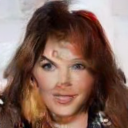} };
                \node[above=13pt of A0, anchor=north] { {\footnotesize \textbf{Pebble}} };
                \node[above=13pt of A4, anchor=north] { {\footnotesize \textbf{CelebA}} };
                \draw[->] ($(A1.north west) + (0, 5pt)$) -- ($(A4.north west) + (0, 5pt)$);
            \end{scope}
        \end{scope}
    \end{tikzpicture}
    \vspace{-6mm}
    \caption{\label{fig:res_nongaussian}
    \textbf{Mapping arbitrary image densities.} We use the same experimental set up as in Figure~\ref{fig:res_cmp_ours_ddim} except that we also try uniform noise, bi-Gaussian noise and a pebble-texture dataset for $p_0$.}
    \vspace{-0mm}
\end{figure}

\begin{figure}[!h]
    \hspace{-5pt}
    \begin{tikzpicture}
        \begin{scope}[shift={(0.0, 0.0\linewidth)}]
            \begin{scope}
                \node[inner sep=0pt,                 ] (A0) { \includegraphics[width=0.2\linewidth, frame]{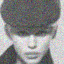} };
                \node[inner sep=0pt, right=-1pt of A0] (A1) { \includegraphics[width=0.2\linewidth, frame]{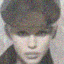} };
                \node[inner sep=0pt, right=-1pt of A1] (A2) { \includegraphics[width=0.2\linewidth, frame]{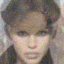} };
                \node[inner sep=0pt, right=-1pt of A2] (A3) { \includegraphics[width=0.2\linewidth, frame]{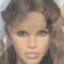} };
                \node[inner sep=0pt, right=-1pt of A3] (A4) { \includegraphics[width=0.2\linewidth, frame]{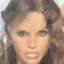} };
                \node[above=13pt of A0, anchor=north] { {\footnotesize \textbf{CelebA gray}} };
                \node[above=13pt of A4, anchor=north] { {\footnotesize \textbf{CelebA color}} };
                \draw[->] ($(A1.north west) + (0, 5pt)$) -- ($(A4.north west) + (0, 5pt)$);
            \end{scope}
        \end{scope}
        \begin{scope}[shift={(0, -0.22\linewidth)}]
            \begin{scope}
                \node[inner sep=0pt,                 ] (A0) { \includegraphics[width=0.2\linewidth, frame]{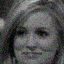} };
                \node[inner sep=0pt, right=-1pt of A0] (A1) { \includegraphics[width=0.2\linewidth, frame]{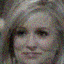} };
                \node[inner sep=0pt, right=-1pt of A1] (A2) { \includegraphics[width=0.2\linewidth, frame]{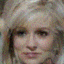} };
                \node[inner sep=0pt, right=-1pt of A2] (A3) { \includegraphics[width=0.2\linewidth, frame]{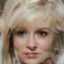} };
                \node[inner sep=0pt, right=-1pt of A3] (A4) { \includegraphics[width=0.2\linewidth, frame]{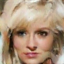} };
            \end{scope}
        \end{scope}
    \end{tikzpicture}
    \vspace{-6mm}
    \caption{\label{fig:restoration}\textbf{Image restoration with IADB.} In this experiment, we use IADB to map grayscale images to color images. 
    The mapping creates a clean image but it does not match the grayscale one. 
    }
    \vspace{-0mm}
\end{figure}

\begin{figure}[!h]
    \centering
    \begin{tikzpicture}
        \begin{scope}[shift={(0.0, 0.0)}]
            \node[inner sep=0pt,                 ] (A0) { \includegraphics[width=0.2\linewidth, frame]{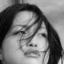} };
            \node[inner sep=0pt, right=-1pt of A0] (A1) { \includegraphics[width=0.2\linewidth, frame]{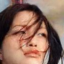} };
            \node[inner sep=0pt, right=-1pt of A1] (A2) { \includegraphics[width=0.2\linewidth, frame]{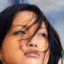} };
            \node[inner sep=0pt, right=-1pt of A2] (A3) { \includegraphics[width=0.2\linewidth, frame]{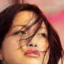} };
            \node[inner sep=0pt, right=-1pt of A3] (A4) { \includegraphics[width=0.2\linewidth, frame]{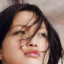} };
            \node[above=13pt of A0, anchor=north] { {\footnotesize condition} };
            \node[above=18pt of A2, anchor=north] { {\footnotesize generations} };
            \draw[|-|] ($(A1.north west) + (0, 5pt)$) -- ($(A4.north east) + (0, 5pt)$);
        \end{scope}        
        \begin{scope}[shift={(0.0, -0.22\linewidth)}]
            \node[inner sep=0pt,                 ] (A0) { \includegraphics[width=0.2\linewidth, frame]{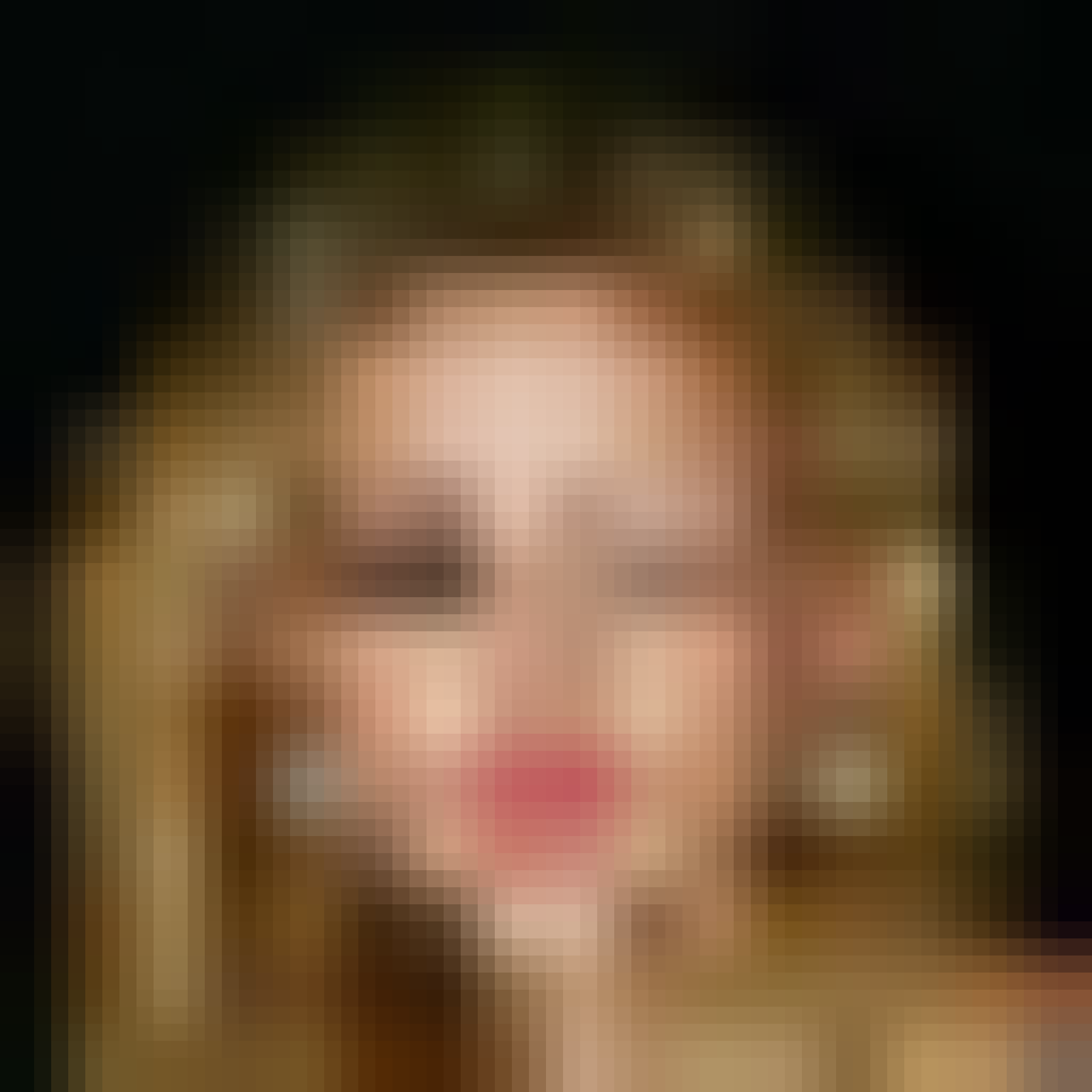} };
            \node[inner sep=0pt, right=-1pt of A0] (A1) { \includegraphics[width=0.2\linewidth, frame]{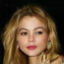} };
            \node[inner sep=0pt, right=-1pt of A1] (A2) { \includegraphics[width=0.2\linewidth, frame]{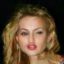} };
            \node[inner sep=0pt, right=-1pt of A2] (A3) { \includegraphics[width=0.2\linewidth, frame]{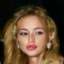} };
            \node[inner sep=0pt, right=-1pt of A3] (A4) { \includegraphics[width=0.2\linewidth, frame]{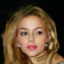} };
        \end{scope}
    \end{tikzpicture}
    \vspace{-6mm}
    \caption{\label{fig:res_conditional}\textbf{Conditional image restoration with IADB.} 
    From a corrupt image (the condition), either decolorization (top) or downscaling (down), we create various restorations using different Gaussian noises $x_0$.
    }
    \vspace{-20mm}
\end{figure}

\clearpage

\section{Convergence of Iterative $\alpha$-(de)Blending}
\label{appendix:b}

\subsection{Preliminaries}

We first recall some properties that are required in the derivation of the limit of Algorithm~1.

\paragraph{The posterior distributions have finite variance.}

The theorem requires that $p_0$ and $p_1$ are of finite variance, such that their respective posterior densities $p_{0|(x, \alpha)}$ and $p_{1|(x, \alpha)}$ are also of finite variance. 
This is because the idea of the proof is that averaging many small random steps (provided by the posterior distributions) converges toward their expectations and it is true only if their variance is finite. 
We use this between Equation~(\ref{eq:proof_algo1_double_limit}) and Equation~(\ref{eq:proof_algo1_expectation_over_interval}).

\paragraph{The expectations of the posterior distributions are continuous.}

If $p_0$ and $p_1$ are classic Riemann-integrable densities, then they are continuous almost everywhere.
Since the blended distributions are essentially convolutions of $p_0$ and $p_1$, it follows that the posterior densities $p_{0|(x, \alpha)}$ and $p_{1|(x, \alpha)}$ are also continuous almost everywhere, and the expectation of their samples $x_{0|(x, \alpha)} \sim p_{0|(x, \alpha)}$ and $x_{1|(x, \alpha)} \sim p_{1|(x, \alpha)}$ are continuous everywhere (the expectation cancels out the null set where they are not continuous).
In summary, for any $x\in\mathbb{R}^d$ and $\alpha \in [0,1]$ we have:
\begin{align}
    \lim_{x' \rightarrow x}~\mathop{\mathbb{E}}\left[x_{0 | (x', \alpha)} \right] = \mathop{\mathbb{E}}\left[x_{0 | (x, \alpha)} \right] ,~~~~
    &\lim_{x' \rightarrow x}~\mathop{\mathbb{E}}\left[x_{1 | (x', \alpha)} \right] = \mathop{\mathbb{E}}\left[x_{1 | (x, \alpha)} \right], \\
    \lim_{\alpha' \rightarrow \alpha}~\mathop{\mathbb{E}}\left[x_{0 | (x, \alpha')} \right] = \mathop{\mathbb{E}}\left[x_{0 | (x, \alpha)} \right] ,~~~~
    &\lim_{\alpha' \rightarrow \alpha}~\mathop{\mathbb{E}}\left[x_{1 | (x, \alpha)} \right] = \mathop{\mathbb{E}}\left[x_{1 | (x, \alpha)} \right].
\end{align}
We use this between Equation~(\ref{eq:proof_algo1_expectation_over_interval}) and Equation~(\ref{eq:proof_algo1}).

\subsection{Objective of the proof}

To prove that Algorithm~1 and Algorithm~2 converge toward the same limit as the number of steps $T$ increases, we need to show that the trajectories of the samples are the same.
This is the case if, in the limit, the derivatives $\frac{\mathrm{d}x_\alpha}{\mathrm{d}\alpha}$ are the same with both algorithms.
The discrete update at step $t$ is:
\begin{align}
\Delta\alpha_t &= \alpha_{t+1} - \alpha_{t} = \frac{1}{T}, \\
\Delta x_{\alpha_t} &= x_{\alpha_{t+1}} - x_{\alpha_{t}},
\end{align}
and we want to prove that for any $\alpha \in [0,1]$ and at point $x_\alpha \in \mathbb{R}^d$ the continuous limit exists and is the same with both algorithms: 
\begin{align}
\frac{\mathrm{d}x_\alpha}{\mathrm{d}\alpha}
=
\lim_{\Delta \alpha \rightarrow 0} \frac{\Delta x_{\alpha}}{\Delta \alpha}.
\end{align}

\subsection{Limit of Algorithm~2.}

In step $t$ of Algorithm~2, we use the average of the posterior samples that are such that
\begin{align}
\label{eq:algo2_posterior_t}
x_{\alpha_{t}} &= 
\left(1-\alpha_{t}\right) \bar{x}_{0|(x_{\alpha_t}, \alpha_t)} + \alpha_{t} \, \bar{x}_{1|(x_{\alpha_t}, \alpha_t)}, \\
\label{eq:algo2_posterior_tplus1}
x_{\alpha_{t+1}} &= 
\left(1-\alpha_{t+1}\right) \bar{x}_{0|(x_{\alpha_t}, \alpha_t)} + \alpha_{t+1} \, \bar{x}_{1|(x_{\alpha_t}, \alpha_t)},
\end{align}
where Equation~(\ref{eq:algo2_posterior_t}) is a property of the average posteriors of $x_{\alpha_t}$ and Equation~(\ref{eq:algo2_posterior_tplus1}) is true by definition in Algorithm~2.
We thus have the discrete difference
\begin{align}
\Delta x_{\alpha_t}
= x_{\alpha_{t+1}} - x_{\alpha_{t}}
= \Delta \alpha_t \, \left(\bar{x}_{1 | (x_{\alpha_t}, \alpha_t)} - \bar{x}_{0 | (x_{\alpha_t}, \alpha_t)}\right).
\end{align}
We obtain the discrete ratio
\begin{align}
\frac{\Delta x_{\alpha}}{\Delta \alpha}
= \bar{x}_{1 | (x_{\alpha}, \alpha)} - \bar{x}_{0 | (x_{\alpha}, \alpha)},
\end{align}
which is independent of $\Delta \alpha$. 
The limit hence exists and is defined by
\begin{align}
\label{eq:proof_algo2}
\boxed{
\frac{\mathrm{d}x_\alpha}{\mathrm{d}\alpha} 
= \lim_{\Delta \alpha \rightarrow 0} \frac{\Delta x_{\alpha}}{\Delta \alpha}
= \frac{\Delta x_{\alpha}}{\Delta \alpha}
= \bar{x}_{1 | (x_{\alpha}, \alpha)} - \bar{x}_{0 | (x_{\alpha}, \alpha)}.
}
\end{align}

\subsection{Limit of Algorithm~1.}

In step $t$ of Algorithm~1,
we sample random posterior samples $x_{0|(x_{\alpha_t}, \alpha_t)}$ and $x_{1|(x_{\alpha_t}, \alpha_t)}$ that are such that
\begin{align}
\label{eq:algo1_posterior_t}
x_{\alpha_{t}} &= \left(1-\alpha_{t}\right) x_{0|(x_{\alpha_t}, \alpha_t)} + \alpha_{t} \, x_{1|(x_{\alpha_t}, \alpha_t)}, \\
\label{eq:algo1_posterior_tplus1}
x_{\alpha_{t+1}} &= \left(1-\alpha_{t+1}\right) x_{0|(x_{\alpha_t}, \alpha_t)} + \alpha_{t+1} \, x_{1|(x_{\alpha_t}, \alpha_t)}, 
\end{align}
where Equation~(\ref{eq:algo1_posterior_t}) is a property of the posteriors of $x_{\alpha_t}$ and Equation~(\ref{eq:algo1_posterior_tplus1}) is true by definition in Algorithm~1.
We thus have the discrete difference
\begin{align}
\Delta x_{\alpha_t} 
&= x_{\alpha_{t+1}} - x_{\alpha_{t}} 
=\Delta\alpha_t \, \left(x_{1 | (x_{\alpha_t}, \alpha_t)} - x_{0 | (x_{\alpha_t}, \alpha_t)} \right).
\end{align}
We obtain the discrete difference for any parameter $\alpha \in [0,1]$ and any location $x_\alpha \in \mathbb{R}^d$
\begin{align}
\Delta x_{\alpha} &= \Delta\alpha \, \left(x_{1 | (x_{\alpha}, \alpha)} - x_{0 | (x_{\alpha}, \alpha)} \right).
\end{align}
Furthermore, increasing the number of steps is equivalent to decomposing each step $\Delta\alpha$ into $N$ smaller steps $\Delta\alpha/N$.
We rewrite the discrete difference as
\begin{align}
\Delta x_{\alpha} &= 
\frac{\Delta\alpha}{N}
\,
\sum_{n=0}^{N-1} 
\left(
x_{1|(x_{\alpha + n\Delta\alpha/N}, {\alpha + n\Delta\alpha/N})}
-
x_{0|(x_{\alpha + n\Delta\alpha/N}, {\alpha + n\Delta\alpha/N})}
\right). 
\end{align}
With this modification, if the derivative exists, it is defined by the limit
\begin{align}
\frac{\mathrm{d}x_{\alpha}}{\mathrm{d}\alpha} 
&= \lim_{\Delta \alpha \rightarrow 0} \lim_{N \rightarrow \infty} \frac{\Delta x_{\alpha}}{\Delta \alpha} = \lim_{\Delta \alpha \rightarrow 0} \lim_{N \rightarrow \infty} \nonumber\\
& 
\frac{1}{N}
\,
\sum_{n=0}^{N-1} 
\left(
x_{1|(x_{\alpha + n\Delta\alpha/N}, {\alpha + n\Delta\alpha/N})}
-
x_{0|(x_{\alpha + n\Delta\alpha/N}, {\alpha + n\Delta\alpha/N})}
\right). 
\label{eq:proof_algo1_double_limit}
\end{align}
Thanks to the \textbf{finite-variance} condition of $p_0$ and $p_1$, the normalized average sum converges toward the average of the posterior samples over $\alpha' \in [\alpha, \alpha+\Delta\alpha]$ as $N$ increases.
\begin{align}
\label{eq:proof_algo1_expectation_over_interval}
\frac{\mathrm{d}x_{\alpha}}{\mathrm{d}\alpha} 
&=
\lim_{\Delta \alpha \rightarrow 0} 
 \mathop{\mathbb{E}}_{\alpha' \in [\alpha, \alpha+\Delta\alpha]}\left[x_{1 | (x_{\alpha'}, \alpha')} \right]
 -
\mathop{\mathbb{E}}_{\alpha' \in [\alpha, \alpha+\Delta\alpha]}\left[x_{0 | (x_{\alpha'}, \alpha')} \right].
\end{align}
Finally, because \textbf{the expectations of the posterior densities are continuous}, we obtain that the expectations over $[\alpha, \alpha+\Delta\alpha]$ converge toward the expectation in $\alpha$, such that
\begin{align}
\boxed{
\label{eq:proof_algo1}
\frac{\mathrm{d}x_{\alpha}}{\mathrm{d}\alpha} 
= \mathop{\mathbb{E}}\left[x_{1 | (x_{\alpha}, \alpha)} \right] - \mathop{\mathbb{E}}\left[x_{0 | (x_{\alpha}, \alpha)} \right] 
= \bar{x}_{1 | (x_{\alpha}, \alpha)} - \bar{x}_{0 | (x_{\alpha}, \alpha)}.
}
\end{align}
This is the same result as in Equation~(\ref{eq:proof_algo2}) with Algorithm~2.

\pagebreak
\section{Variant Formulations}
\label{appendix:c}

We derive the variant formulations introduced in Section~4.

\paragraph{Blended samples.}

A blended sample is by definition the blending of its posterior samples
\begin{align}
x_{\alpha_{t}} &= (1-\alpha_t) \, x_0 + \alpha_t \, x_1.
\end{align}
Since blending is linear, a blended sample is also the blending of the average of its posterior samples:
\begin{align}
x_{\alpha_{t}} &= (1-\alpha_t) \, \bar{x}_0 + \alpha_t \, \bar{x}_1.
\end{align}
We can thus rewrite its average posteriors samples $\bar{x}_0$ and $\bar{x}_1$ in the following way:
\begin{align}
\bar{x}_0 &= \frac{x_{\alpha_{t}}}{1-\alpha_t} - \frac{\alpha_t \, \bar{x}_1}{1-\alpha_t}, \label{eq:appc_x0}\\
\bar{x}_1 &= \frac{x_{\alpha_{t}}}{\alpha_t} - \frac{(1-\alpha_t) \, \bar{x}_0}{\alpha_t}. \label{eq:appc_x1}
\end{align}

\paragraph{Variant (a):}

In the vanilla version of the algorithm, a blended sample of parameter $\alpha_{t+1}$ is obtained by blending $\bar{x}_0$ and $\bar{x}_1$:
\begin{align}
x_{\alpha_{t+1}} &= (1-\alpha_{t+1}) \, \bar{x}_0 + \alpha_{t+1} \, \bar{x}_1. \label{eq:appc_x_alpha_prime}
\end{align}

\paragraph{Variant (b):}

By expanding $\bar{x}_1$ from Equation~(\ref{eq:appc_x_alpha_prime}) using Equation~(\ref{eq:appc_x1}), we obtain
\begin{align}
x_{\alpha_{t+1}} 
&= (1-\alpha_{t+1}) \, \bar{x}_0 + \alpha_{t+1} \, \bar{x}_1, \nonumber\\
&= (1-\alpha_{t+1}) \, \bar{x}_0 + \alpha_{t+1} \, \left(\frac{x_{\alpha_{t}}}{\alpha_t} - \frac{(1-\alpha_t) \, \bar{x}_0}{\alpha_t}\right), \nonumber\\
&=  \left(1- \alpha_{t+1} - \frac{\alpha_{t+1} \,(1-\alpha_t)}{\alpha_t}\right) \, \bar{x}_0 + \frac{\alpha_{t+1}}{\alpha_t} \, x_{\alpha_{t}} ,\nonumber\\
&=  \left(1- \frac{\alpha_{t+1}}{\alpha_t}\right) \, \bar{x}_0 + \frac{\alpha_{t+1}}{\alpha_t} \, x_{\alpha_{t}} ,\nonumber\\
&= \bar{x}_0 + \frac{\alpha_{t+1}}{\alpha_t} \left(x_{\alpha_{t}} - \bar{x}_0 \right).
\end{align}

\paragraph{Variant (c):}

By expanding $\bar{x}_0$ from Equation~(\ref{eq:appc_x_alpha_prime}) using Equation~(\ref{eq:appc_x0}), we obtain
\begin{align}
x_{\alpha_{t+1}} 
&= (1-\alpha_{t+1}) \, \bar{x}_0 + \alpha_{t+1} \, \bar{x}_1, \nonumber\\
&= (1-\alpha_{t+1}) \,\left(\frac{x_{\alpha_{t}}}{1-\alpha_t} - \frac{\alpha_t \, \bar{x}_1}{1-\alpha_t}\right) + \alpha_{t+1} \, \bar{x}_1, \nonumber\\
&= \left( \alpha_{t+1} - \frac{(1-\alpha_{t+1}) \, \alpha_t}{1-\alpha_t}\right) \, \bar{x}_1 + \frac{1-\alpha_{t+1}}{1-\alpha_t} \, x_{\alpha_{t}} , \nonumber\\
&= \left( 1 - \frac{1-\alpha_{t+1}}{1-\alpha_t}\right) \, \bar{x}_1 + \frac{1-\alpha_{t+1}}{1-\alpha_t} \, x_{\alpha_{t}} , \nonumber\\
&= \bar{x}_1 + \frac{1-\alpha_{t+1}}{1-\alpha_t} \, \left(x_{\alpha_{t}} - \bar{x}_1\right). 
\end{align}

\paragraph{Variant (d):}

By rewriting $\alpha_{t+1}=\alpha_{t+1}+\alpha_t-\alpha_t$ in the definition of $x_{\alpha_{t+1}}$, we obtain
\begin{align}
x_{\alpha_{t+1}} 
&= (1-\alpha_{t+1}) \, \bar{x}_0 + \alpha_{t+1} \, \bar{x}_1, \nonumber\\
&= (1-\alpha_{t+1} + \alpha_t-\alpha_t) \, \bar{x}_0 + \left(\alpha_{t+1}+\alpha_t-\alpha_t\right) \, \bar{x}_1, \nonumber\\
&= (1-\alpha_t) \, \bar{x}_0 + \alpha_t \, \bar{x}_1 + \left(\alpha_{t+1}-\alpha_t\right) \, \left(\bar{x}_1 - \bar{x}_0\right), \nonumber\\
&= x_{\alpha_{t}} + \left(\alpha_{t+1}-\alpha_t\right) \, \left(\bar{x}_1 - \bar{x}_0\right).
\end{align}
\pagebreak
\section{Relation to DDIM}
\label{appendix:d}

\begin{figure}[!h]
    \centering
    {\small
    \begin{tikzpicture}
        \node[inner sep=0] (B0) { \includegraphics[width=0.32\linewidth]{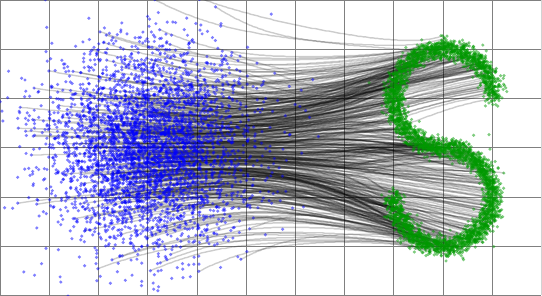} };
        \node[inner sep=0, right=2pt of B0] (B1) { \includegraphics[width=0.32\linewidth]{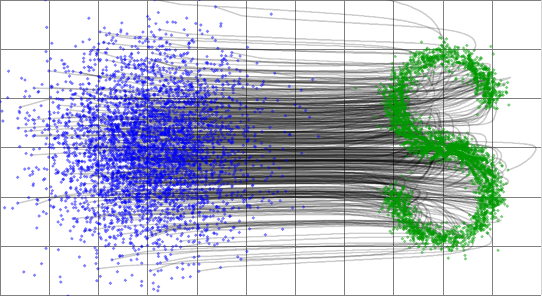} };
        \node[inner sep=0, right=2pt of B1] (B2) { \includegraphics[width=0.32\linewidth]{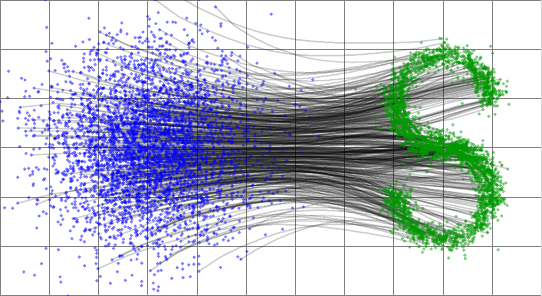} };

        \node[inner sep=0, above=10pt of B0, anchor=north] {a) IADB};    
        \node[inner sep=0, above=10pt of B1, anchor=north] {b) DDIM};    
        \node[inner sep=0, above=10pt of B2, anchor=north] {c) DDIM rescaled (Equation~\ref{eqn:appendixE_xt})};    
    \end{tikzpicture}
    }
\vspace{-6mm}
\caption{\label{fig:appendix_rescaled_ddim}
We trained an MLP with 5 hidden layers of 64 filters to learn $D_\theta$ for IADB (a) and the same architecture to learn $\epsilon_\theta$ for DDIM~(b) and~(c). For (c), we convert points generated by DDIM using the scaling equation. The trajectories of the samples for IADB (a) and DDIM rescaled (c) match.
}
\end{figure}

In this section, we follow the notation of~\cite{song2021ddim}: $x_0$ is a sample of a target density and $\epsilon$ is a random Gaussian sample.
The denoiser of DDIM is defined such that, for an input $x_t = \sqrt{\alpha_t} x_0 + \sqrt{1-\alpha_t} \epsilon$, it learns $\epsilon^{(t)} \left( x_t \right) = \bar{\epsilon}$.
We define
\begin{align}
    y_t &= \dfrac{x_t}{\sqrt{\alpha_t} + \sqrt{1-\alpha_t}} = \beta_t x_0 + \left( 1 - \beta_t \right) \epsilon,
\end{align}
where $\beta_t = \dfrac{\sqrt{\alpha_t}}{\sqrt{\alpha_t} + \sqrt{1-\alpha_t}}$. $y_t$ is an alpha-blended sample such as the one we defined in Section~3.
It follows that we have
\begin{align}
    \dfrac{{x}_t}{\sqrt{\alpha_t}} &= \dfrac{y_t}{\beta_t}.
    \label{eqn:appendixE_xt}
\end{align}
We now turn to Equation~(13) of~\cite{song2021ddim}:
\begin{align}
    \dfrac{x_{t+1}}{\sqrt{\alpha_{t+1}}} &= \dfrac{x_t}{\sqrt{\alpha_t}} + \left(\sqrt{\dfrac{1 - \alpha_{t+1}}{\alpha_{t+1}}} - \sqrt{\dfrac{1-\alpha_t}{\alpha_t}} \right) \epsilon^{(t)}\left(x_t\right),
\end{align}
By injecting the scaled coordinate from Equation~(\ref{eqn:appendixE_xt}) into this expression, we obtain:
\begin{align}
    \dfrac{y_{t+1}}{\beta_{t+1}} &= \dfrac{y_t}{\beta_t} + \left(\sqrt{\dfrac{1 - \alpha_{t+1}}{\alpha_{t+1}}} - \sqrt{\dfrac{1-\alpha_t}{\alpha_t}} \right) \bar{\epsilon} \nonumber\\
\Rightarrow 
    y_{t+1} 
     &= y_t \dfrac{\beta_{t+1}}{\beta_t} + \dfrac{1}{\beta_t} {\beta_t \beta_{t+1}} \left(\sqrt{\dfrac{1 - \alpha_{t+1}}{\alpha_{t+1}}} - \sqrt{\dfrac{1-\alpha_t}{\alpha_t}} \right) \bar{\epsilon}.
\label{eqn:appendixE_cpx}
\end{align}
Since
\begin{align}
    \beta_{t+1}\beta_t \left(\sqrt{\dfrac{1 - \alpha_{t+1}}{\alpha_{t+1}}} - \sqrt{\dfrac{1-\alpha_t}{\alpha_t}} \right) &= \beta_{t} \left(1-\beta_{t+1}\right) - \left( 1 - \beta_{t} \right) \beta_{t+1}, \nonumber \\
    &= \beta_t - \beta_{t+1}
\end{align}
we can simplify Equation~(\ref{eqn:appendixE_cpx}) to
\begin{align}
    y_{t+1} = y_t \dfrac{\beta_{t+1}}{\beta_t} + \dfrac{\beta_t - \beta_{t+1}}{\beta_t} \bar{\epsilon}
     &= \bar{\epsilon} + y_t \dfrac{\beta_{t+1}}{\beta_t}  - \dfrac{\beta_{t+1}}{\beta_t}\bar{\epsilon}, \nonumber\\
     &= \bar{\epsilon} + \dfrac{\beta_{t+1}}{\beta_t} \left( y_t - \bar{\epsilon}\right).
     \label{eqn:appendixE_final}
\end{align}
This last form is exactly variant-(b) of IADB (see Table~1).
We confirm this experimentally in Figure~\ref{fig:appendix_rescaled_ddim}.

\end{document}